\documentclass[10pt,journal,compsoc]{IEEEtran}

\usepackage{tabu}                      
\usepackage{booktabs}                  
\usepackage{lipsum}                    
\usepackage{mwe}                       
\usepackage{multirow}
\usepackage{makecell}
\usepackage{enumitem}
\usepackage{verbatim}
\usepackage[normalem]{ulem}
\usepackage{xspace}
\usepackage{colortbl}
\usepackage{color}
\usepackage{xcolor}
\usepackage{ulem}
\usepackage{xurl}
\usepackage{amsmath,amsfonts}
\usepackage{algorithmic}
\usepackage{algorithm}
\usepackage{array}
\usepackage[caption=false,font=normalsize,labelfont=sf,textfont=sf]{subfig}
\usepackage{textcomp}
\usepackage{stfloats}
\usepackage{url}
\usepackage{verbatim}
\usepackage{graphicx}
\usepackage{cite}
\usepackage[numbers,sort&compress]{natbib}
\usepackage{hyperref}
\hypersetup{
    colorlinks=true,
    linkcolor=blue,
    filecolor=magenta,      
    urlcolor=cyan,
    pdftitle={Overleaf Example},
    pdfpagemode=FullScreen,
    }
\usepackage{wrapfig}

\usepackage{mathptmx}                  

\begin{document}

\renewcommand{\arraystretch}{1.5}
\newcommand{\eg}{\emph{e.g.}}
\newcommand{\ie}{\emph{i.e.}}
\newcommand{\etc}{\textit{etc}}
\newcommand{\q}[1]{\textit{``#1''}}
\newcommand{\tool}{\emph{FinFlier}\xspace}
\newcommand*{\img}[1]{%
    \makebox[\baselineskip]{ 
    \raisebox{-.2\baselineskip}{ 
        \includegraphics[
          height=\baselineskip,
          width=\baselineskip,
          keepaspectratio,
          trim=0 0 10 0,clip
        ]{#1}
    }
    }
}

\title{FinFlier: Automating Graphical Overlays for Financial Visualizations with Knowledge-Grounding Large Language Model}

\author{Jianing Hao, Manling Yang, Qing Shi, Yuzhe Jiang, Guang Zhang, Wei Zeng,~\IEEEmembership{Member,~IEEE}
\thanks{J. Hao, L. Yang, Q. Shi, G. Zhang and W. Zeng are with the Hong Kong University of Science and Technology (Guangzhou). G. Zhang and W. Zeng are also with the Hong Kong University of Science and Technology Email: \{jhao768@connect, myang838@connect, brantshi@, guangzhang@, weizeng@\}.hkust-gz.edu.cn}%
\thanks{Y. Jiang is with the Hong Kong University of Science and Technology. Email: yjiangbu@connect.ust.hk.}%
\thanks{W. Zeng is the corresponding author.}}

\markboth{submitted to IEEE Transactions on Visualization and Computer Graphics}%
{Hao \MakeLowercase{\textit{et al.}}: A Sample Article Using IEEEtran.cls for IEEE Journals}


\maketitle

\begin{abstract}
Graphical overlays that layer visual elements onto charts, are effective to convey insights and context in financial narrative visualizations.
However, automating graphical overlays is challenging due to complex narrative structures and limited understanding of effective overlays. 
To address the challenge, we first summarize the commonly used graphical overlays and narrative structures, and the proper correspondence between them in financial narrative visualizations, elected by a survey of 1752 layered charts with corresponding narratives.
We then design \tool, a two-stage innovative system leveraging a knowledge-grounding large language model to automate graphical overlays for financial visualizations.
The \emph{text-data binding} module enhances the connection between financial vocabulary and tabular data through advanced prompt engineering, and the \emph{graphics overlaying} module generates effective overlays with narrative sequencing.
We demonstrate the feasibility and expressiveness of \tool through a gallery of graphical overlays covering diverse financial narrative visualizations. 
Performance evaluations and user studies further confirm system's effectiveness and the quality of generated layered charts.
\end{abstract}

\begin{IEEEkeywords}
Graphical overlay, financial narrative, LLM.
\end{IEEEkeywords}

\section{Introduction}
\begin{figure*}
    \centering
    \includegraphics[width=0.98\linewidth]{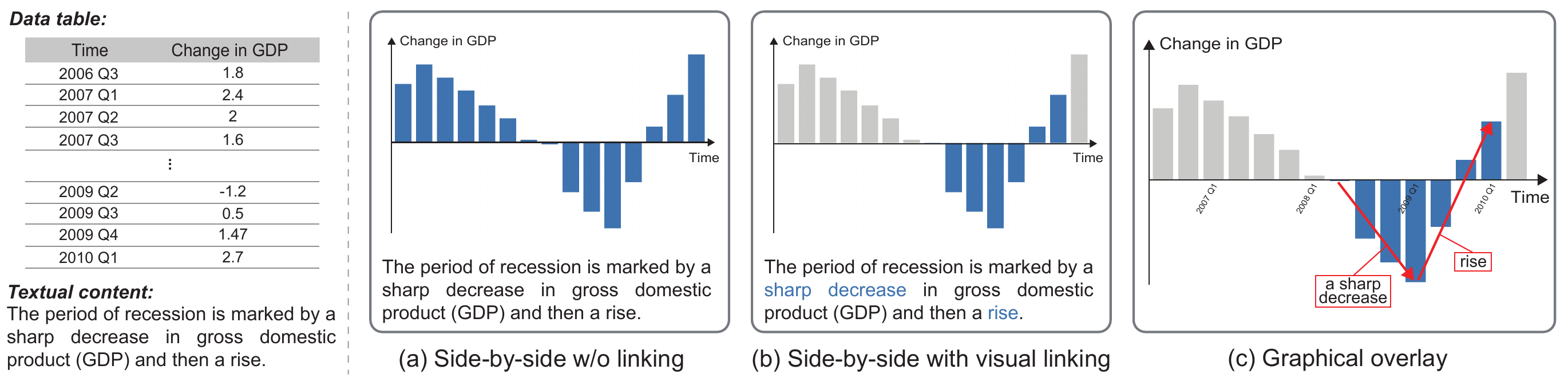}
    \vspace{-2mm}
    \caption{The narrative introduces the change in GDP growth during the 2008 Great Recession. (a) displays the side-by-side interplay without visual linking, (b) shows the side-by-side interplay with visual linking, and (c) utilizes graphical overlays.}
    \vspace{-3mm}
    \label{fig:intro}
\end{figure*}

\IEEEPARstart{F}{inancial} narratives are frequently conveyed through textual descriptions and data tables that can be rendered as charts.
Reading financial narratives entails a balance between swimming in extensive textual content and extracting valuable insights from the charts~\cite{tellingfinance_2000}.
Proficient reading in this context proves particularly advantageous for stakeholders in pursuit of in-depth insights, including stock traders and marketing managers~\cite{narrineco_2024}.
However, this task is challenging, as readers must decode intricate narrative structures within financial descriptions and associate them with complex data patterns~\cite{tellingfinance_2000}.
Consider a scenario where a reader is engrossed in an article that discusses recession periods characterized by a significant decrease in gross domestic product (GDP) followed by a subsequent increase.
This information is presented in both textual descriptions and a data table, as shown in Figure~\ref{fig:intro} (left).
Conventionally, the data table is converted into a chart, and the textual descriptions are placed side-by-side (Figure~\ref{fig:intro} (a)). 
This side-by-side placement requires the reader to mentally connect the narrative with the changing patterns in the chart, resulting in cognitive load as they attempt to decipher the relationship between the narratives and the visual representation~\cite{narrvis_2010}.

Visualization-text interplay, which bridges textual descriptions with data tables or charts, has emerged as a pivotal concept in the realm of narratives~\cite{lai2020automatic, pinheiro2022charttext, chen2022crossdata, masson2023charagraph, DataTales_2023}.
As depicted in Figure~\ref{fig:intro} (b), phrases \q{sharp decrease} and \q{rise} are highlighted, with corresponding bars highlighted in blue.
Nonetheless, this method is restricted by its reliance on a limited set of visual cues for effective association.
Textual descriptions primarily serve for highlighting, while financial narratives and data patterns are often intricate and multifaceted~\cite{narrineco_2024}, extending beyond basic highlights.
Alternatively, graphical overlays that layer visual elements onto charts, can effectively convey insights~\cite{graphicalo_2012} and add context information~\cite{hullman2013contextifier} for finance narratives.
For example, in Figure~\ref{fig:intro} (c), the decrease and rise are intuitively represented with downward and upward arrows, complemented by accompanying texts placed alongside.

There is a growing interest in embracing graphical overlays for financial visualizations.
Nevertheless, automating the creation of graphical overlays presents challenges due to the vast design possibilities for chart layering patterns and the intricate narrative structures in financial text descriptions.
While design space and practices of graphical overlays have been explored (\eg,~\cite{srinivasan2018augmenting, godesign_2023, hullman2013contextifier}), effective layering patterns for conveying financial narratives are not thoroughly examined.
The complexity increases when considering different chart types and matching diverse financial narrative structures.
Moreover, the textual descriptions prevalent in financial narratives, often exhibit complex structures and specialized vocabularies~\cite{finance_2004}. 
Comprehending such text requires familiarity with domain-specific terminology and data context~\cite{visltr_2024}.
For instance, establishing the association between the phrase \q{sharp decrease} and underlying data patterns necessitates the understanding of what types of data declines are deemed as \emph{`sharp'} within the context of finance.
Traditional natural language processing (NLP) approaches rely heavily on substantial amounts of labeled training data~\cite{zhou2023one}, which is resource-scarce within the financial domain.
Furthermore, there is a demand for authoring tools that complement automated approaches with support for customized refinement of layering designs~\cite{linktv_2024}. 

In this paper, we propose \tool, a two-stage innovative system leveraging a knowledge-grounding large language model (LLM) to automate graphical overlays for financial visualizations.
The design of \tool is informed by a comprehensive design space of proper correspondence between financial narratives and graphical overlays, which is grounded in insights gained from practical utilization of graphical overlays and financial narrative visualizations (Sect.~\ref{sec:space}).
On the basis, we build two essential modules in \tool. 
First, the \emph{text-data binding} module (Sect.~\ref{ssec:connection}) leverages a knowledge-grounding LLM, utilizing a set of prompt engineering techniques to optimize the identification of complex structures in financial narratives.
Second, the \emph{graphics overlaying} module (Sect.~\ref{ssec:implementation}) automatically associates derived narrative structure with appropriate graphical overlays.
We also develop an interactive visual interface to complement the automated approach, allowing users to generate and configure layered charts with personalized design preferences (Sect.~\ref{ssec:interface}).
We elaborate on the feasibility and expressiveness of \tool via examples that encompass a wide range of visual and textual narratives (Sect.~\ref{ssec:case}). 
Experimental results underscore the capability of \tool to facilitate a rapid understanding of key information contained in the text (Sect.~\ref{ssec:eva_txt-data}), and foster a more intuitive and informed interaction with both textual and visual content (Sect.~\ref{ssec:user_study}).
The contributions of this work are summarized as follows.
\begin{itemize}
    \item \textbf{Correspondence between graphical overlays and financial narrative visualizations}. Through the survey of 1752 layered charts from various financial and academic sources, we identify commonly used graphical overlays.
    We also examine common narrative structures within financial textual content, and construct a correspondence between graphical overlays and financial narratives.
    The corpus of 1752 layered charts, labeled with their sources, categories, and statistical information, is available at \url{https://osf.io/xb2rd/}.
    
    \item \textbf{System construction.} We design a novel system \tool to automate graphical overlays for financial narrative visualizations. 
    \tool comprises two modules: the \emph{text-data binding} module improves the connection between financial vocabulary and tabular data using a financial knowledge-grounding LLM, and the \emph{graphics overlaying} module generates layered charts considering the proper correspondence between graphical overlays and financial narratives.

    \item \textbf{Evaluation.} We conduct case studies, quantitative experiments, as well as qualitative experiments to validate the feasibility and effectiveness of \tool. 
    These assessments demonstrate the system's capability to achieve accurate text-data binding and generate high-quality layered charts.
\end{itemize}
\section{Related Work}
\subsection{Financial Narrative Visualization}
Narrative visualization combines data visualization with storytelling to communicate complex information in an engaging and interactive manner~\cite{narrvis_2010,segel2010narrative}.
The approach encompasses a range of genres, such as annotated charts~\cite{graphicalo_2012}, flow charts~\cite{calliope_2020}, and animations~\cite{dataplayer_2023}.
Given the importance of conveying information within the finance, many authoring tools have been developed to create effective financial narrative visualizations.
For example, Contextifier~\cite{hullman2013contextifier} automatically generates customized, annotated stock behavior visualizations based on news articles, making it easier for readers to understand the impact of market news on stock prices.
Similarly, DeepClue~\cite{shi2018deepclue} identifies key factors in stock price prediction models and supports users in analyzing stock market by providing interactive, layered visualizations.

Nevertheless, designing effective financial narrative visualizations remains challenging, as it involves conveying data stories about financial events, trends, or phenomena.
Specifically, data in financial domain can be categorized into stocks, funds, economic indicators, trading, risk, and company information~\cite{ko_2016_survey}.
These data typically exhibit multifaceted characteristics such as time-series, cross-sectional, and tabular aspects~\cite{koop2022analysis}, promoting various analysis types including horizontal, vertical, and combined analyses.
Moreover, in financial narratives like annual reports, specialized terminologies are commonly employed to articulate these patterns.
For instance, the term `\emph{bullish trend}' may be used to denote an increasing trend in the stock market~\cite{brown2020language}.
When reading such textual descriptions, people must comprehend the terminology and connect it to the underlying data.
This is non-trivial, as readers need to understand complex narrative structures and there can be vague references to the context and data.
To address these challenges, this work seeks to develop an automated approach for seamless integration of financial narratives with layered charts.
Our contribution entails a comprehension of the design space for financial narrative visualizations and introducing a knowledge-grounding LLM-based approach to accomplish the goal.

\subsection{Visualization-text Interplay}
Recent studies have highlighted the cognitive challenges associated with the separation of text and charts, which can cause a split-attention effect and increase the cognitive burden on users~\cite{latif2021kori, wordsize_2017}.
In contrast, integrating visualization with text can enhance comprehension by allowing readers to process visual and textual information simultaneously~\cite{storytelling_2013}.
This integration, known as visualization-text interplay, has been studied and shown to facilitate human cognition~\cite{ren2017chartaccent,zhi2019linking,badam_2019_elastic, head2021augmenting}.
Theoretical research has extensively investigated the necessity of applying visual marks~\cite{kong2014extracting}, the balance between visualization and text~\cite{stokes2022striking}, and the design space of visualization-text interplay~\cite{latif2018exploring}.
The studies have promoted the design and development of various authoring tools. 
Lai et al.~\cite{lai2020automatic} developed a system that automatically annotates raster images of bar, pie, and scatter plots based on textual descriptions using deep neural networks. 
Similarly, authoring tools such as Kori~\cite{latif2021kori}, CrossData~\cite{chen2022crossdata}, and Charagraph~\cite{masson2023charagraph} establish links between text and visualizations to aid in creating data documents and generating charts for data-rich paragraphs.
However, these tools typically place text and visualization side by side, often without any visual cues or with colors connecting them.
This requires effort to associate corresponding components in the text and visualization.

Graphical overlay techniques, which layer visual elements onto charts, offer another method of visualization-text interplay~\cite{graphicalo_2012}.
Applying appropriate graphical overlays can provide clear visual guidance to communicate key concepts without confusion~\cite{kong2017internal}.
For instance, Kong et al.~\cite{kong2017internal} recommended adding complementary elements such as text, arrows, rectangles, and brackets to emphasize the specific areas of charts.
Recently, LTV~\cite{linktv_2024} enables reformatting the styles of graphical overlays, to support flexible customization of text-visualization links.
However, automatically applying graphical overlays to financial visualizations is challenging due to the intricate narrative structures and specific design needs of the finance domain.
To address this gap, we first conduct a preliminary survey to distill the design space of graphical overlays tailored for financial visualizations.
Our findings cover diverse layering patterns and strategies for conveying complex financial narratives effectively.
We further enhance the automatic linking between financial narratives and visualizations, by leveraging the power of a knowledge-grounding LLM to overcome the limitations of traditional NLP techniques.

\subsection{LLM for Text-data Binding}
Heuristic approaches for text-data binding generally follow a two-stage process:
first, extracting meaningful texts using techniques such as word scoring~\cite{badam_2019_elastic}, grammar trees~\cite{chen2022crossdata}, and text reference extractor~\cite{emphchecker_2023}; 
and second, matching text with data using rule-~\cite{kim2018facilitating} or keyword-based~\cite{badam_2019_elastic} matching, as well as feature-word-topic model~\cite{difference_2023}.
However, these rule-based methods are limited to specialized vocabularies and narrative structures~\cite{finance_2004}.
Deep learning has also shown success in parsing textual descriptions for further connecting data items in charts~\cite{lai2020automatic, linktv_2024}. 
Nonetheless, the learning-based methods depend on carefully human-labeled corpora, which require intensive annotators with domain-specific knowledge.
The emergence of LLMs like GPT, has revolutionized NLP by enabling impressive capabilities with proper prompt engineering, such as few-shot in-context learning~\cite{liu2022few}, chain-of-thought (CoT) reasoning~\cite{cot_2022}, and instruction following~\cite{ouyang2022training}. 
With these capabilities, LLMs have significantly improved NLP applications across various specialized domains including finance~\cite{huang2023finbert, yang2023fingpt}.
Advances in LLMs also bring new opportunities for text-data binding, addressing challenges in text extraction and text-data matching~\cite{GAIsur_2024}.
Recently, Data Player~\cite{dataplayer_2023} utilizes LLMs with few-shot learning to form semantic connections between text and visuals, generating high-quality data videos.

However, LLMs with few-shot learning struggle with text-data binding due to the complex structure of data tables and intricate narratives~\cite{chain_of_table}.
To overcome the challenges, we employ a financial knowledge-grounding LLM that integrates domain-specific knowledge into LLMs, including optimization techniques of output constraint, CoT, and dynamic prompt. Experiment results demonstrate the effectiveness of our proposed approach in identifying complex financial narrative structures, and enhancing the connection between financial vocabularies and data tables.

\section{Preliminary Study and Design Space}\label{sec:space}
Studies have summarized the design space for graphical overlays in general~\cite{srinivasan2018augmenting, godesign_2023}.
However, graphical overlays in financial narrative visualization remain under-explored.
To this end, we out to survey the design space of graphical overlays (Sect.~\ref{ssec:graph_overlay}) and narrative structure (Sect.~\ref{ssec:fin_pattern}) within financial narratives.
Through statistics results and discussions with domain experts, we distill a proper correspondence between graphical overlays and financial narratives (Sect.~\ref{ssec:correspondence}).

\begin{figure*}[t]
    \centering
    \includegraphics[width=0.98\textwidth]{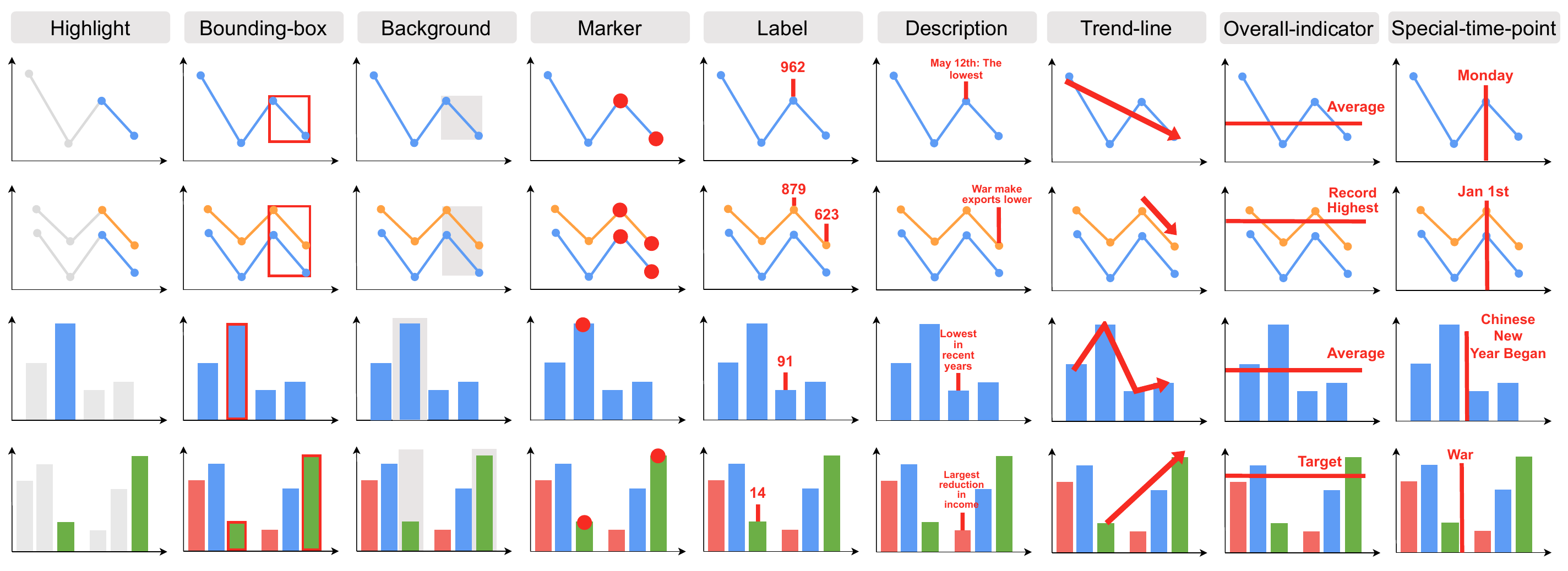}
    \vspace{-4mm}
    \caption{Examples of graphical overlay techniques organized by category. The four rows from top to bottom correspond to the four chart types: single-line chart, multi-line chart, single-bar chart, multi-bar chart, and the nine columns correspond to the nine categories.}
    \vspace{-4mm}
    \label{fig:go_map}
\end{figure*}

\subsection{Graphical Overlays in Financial Visualization}
\label{ssec:graph_overlay}

Graphical overlays are popular in both financial domain and academic publications for improving the overall interpretability and clarity~\cite{latif2021kori, wordsize_2017}.
Here, we focus on four common chart types, \ie, \emph{single-line chart, multi-line chart, single-bar chart}, and \emph{multi-bar chart}.
For clarity, we define visual elements as the constituent parts of a chart, such as line segments in a line chart or bars in a bar chart.
We refer to layered charts as the outcome of layering graphical overlays onto base charts, whereas the base charts are denoted as plain charts without any graphical overlays.

To summarize the design space of graphical overlays in the financial domain, we had several rounds of discussions with a collaborating finance expert with over 10 years of experience.
We decided to use three sources of layered charts: financial newspapers, public marketing reports, and company analysis reports.
For financial newspapers, we collected the layered charts from the Graphic Detail section in \emph{The Economist}\footnote{\url{https://www.economist.com/graphic-detail}} in the recent five years (2018 - 2023), yielding a total of 613 charts.
These layered charts are readily accessible and provide a rich, varied dataset. 
Alternative financial newspapers like Bloomberg, WSJ, and Yahoo Finance lack dedicated sections for layered charts, complicating the filtering and extraction process.
For public marketing reports, we collected 381 layered charts from reports in the Markets section of \emph{Financial Times}\footnote{\url{https://www.ft.com/markets}} (2018 - 2023).
For company analysis reports, we extracted 309 layered charts from 100 \emph{MorningStar}\footnote{\url{https://www.morningstar.com/stocks}} analysis reports on different companies.
These sources encompass a wide range of financial information and presentation styles, aligning with our target scenario.
In addition, these sources cater to a broad audience, encompassing not only financial practitioners but also general users.

To summarize graphical overlay techniques in academic publications, we collected layered charts from professional academic journals and conferences: IEEE TVCG, CGF, PacificVis, and ACM CHI, all from year 2007 to the present.
Given the large number of papers published each year and the breadth of ACM CHI, we selected full papers by searching keywords such as \q{finance}, \q{narrative}, \q{chart}, \etc. including variations like \q{financial}, to identify papers that might include layered charts.
After collecting the papers, we extracted charts that met the definition of layered charts, while omitting visual analytics interfaces with multiple views.
This process yielded a total of 449 layered charts: 22 from 11 PacificVis papers, 77 from 49 ACM CHI papers, 63 from 43 CGF papers, and 287 from 242 TVCG papers.
The inclusion of academic sources allows us to capture the theoretical and methodological advancements in the field, providing a valuable supplement to the more application-focused examples from financial news and reports.

The dual-source approach, encompassing both practical and research perspectives, allows us to construct a more comprehensive understanding of graphical overlays in practice.
In total, we collected 1752 layered charts with corresponding narratives.
Next, we summarize graphical overlays in the following categories, referring to previous studies~\cite{graphicalo_2012, srinivasan2018augmenting}.

\begin{itemize}
    \item \textbf{Highlight}: 
    Recoloring some visual elements in the chart to be more distinctive from others~\cite{latif2018exploring}.
    \item \textbf{Bounding-box}: Adding colored rectangles in regions of some visual elements.
    \item \textbf{Background}: Modifying the background of some visual elements to create visual contrast and enhance the prominence of certain data points.
    \item \textbf{Marker}: Inserting symbols or shapes onto some visual elements.
    
    \item \textbf{Label}: Attaching text labels to some visual elements to provide additional \emph{contextual} information with no more than three words or numbers.
    
    \item \textbf{Description}: Incorporating textual descriptions alongside visual elements to provide \emph{detailed} insights and aid in comprehension. 
    A description typically includes mixed forms of text and data of more than three words, or a complete sentence.

    \item \textbf{Trend-line}: Adding arrow lines to visually depict upward, downward, or stable trends.
    
    \item \textbf{Overall-indicator}: Integrating summary indicators such as mean, maximum and minimum onto the chart to offer an immediate understanding of the overall data.
    \item \textbf{Special-time-point}: Emphasizing specific time points by lines to highlight significant events or occurrences.
\end{itemize}

\begin{figure}[t]
    \centering
    \includegraphics[width=0.9\linewidth]{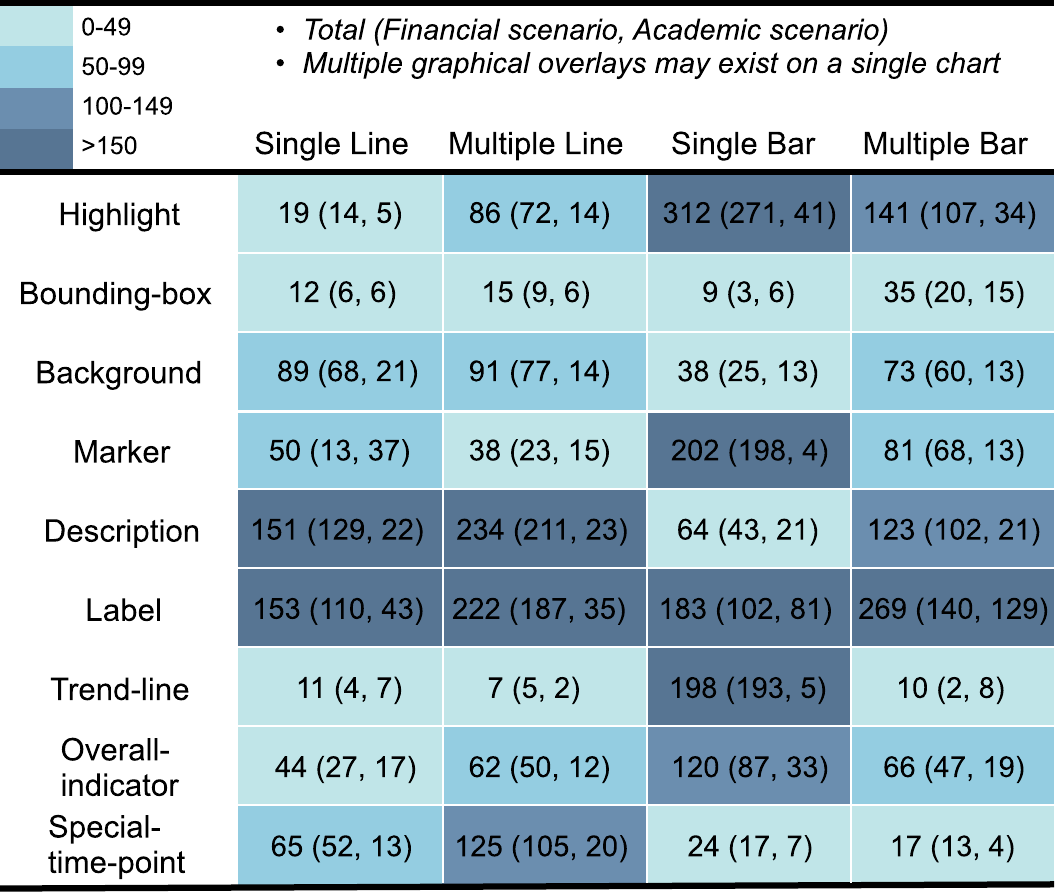}
    \vspace{-4mm}
    \caption{The statistical results of graphical overlays in our collected corpus of layered charts.}
    \vspace{-3mm}
    \label{fig:stat}
\end{figure}

The corresponding visual representations for the four chart types are shown in Figure~\ref{fig:go_map}.
When collecting the data, we found that these techniques often do not appear alone, but are more often used in combination in real-world narrative visualizations.
For example, annotation systems for financial narrative visualizations, like Contextifier~\cite{hullman2013contextifier} and DeepClue~\cite{shi2018deepclue}, leverage combinations of \emph{Marker}, \emph{Description}, \emph{Background}, to promote financial narratives.
Therefore, this categorization splits all overlays in the layered chart into the smallest units, which can better describe the combination of overlays.
After collecting the corpus, we then identified the commonly used overlaying techniques for each chart type, by coding the overlaying technique and counting the number in each of the nine categories.
One of the authors and the collaborating finance expert independently coded all layered charts in the corpus first.
They met over two sessions to post typical layered chart examples and reached a mutual understanding with sufficient operationalization.
They then completed coding the corpus of layered charts and checked how closely their coding matched.
At this point, they computed inter-rater reliability using Cohen's kappa ($\kappa = 0.82$) in order to assess the reliability of the coding results.
They then discussed all the mismatches until they came to an agreement, revising the categories and definitions as needed to reach 100\% consensus.

The statistical results are presented in Figure~\ref{fig:stat}, with the numbers indicating total appearances, and appearances in financial and academic sources, respectively.
For example, for `\emph{Highlight}' \& `\emph{single line chart}' cell, `\emph{19 (14, 5)}' indicates that there are a total of 19 layered single-line charts with the \emph{highlight} technique, with 14 coming from financial sources and 5 from academic papers.
Since multiple graphical overlays may exist simultaneously on a single base chart, a layered chart can be counted in several different overlaying categories. 
According to our statistics, 1139 layered charts contain the combination of multiple graphical overlay techniques, accounting for 65.04\% of the total.

\subsection{Financial Narrative Structure}
\label{ssec:fin_pattern}
To effectively layer graphical overlays on financial visualizations, it is essential to decompose financial narratives into a structured format.
This is crucial for directing readers' attention to specific portions of the textual content~\cite{hyland_1998}.
To better understand the structure, we summarized typical financial narrative examples that can be categorized into three types of analysis: \emph{horizontal analysis}, \emph{vertical analysis} and \emph{combined analysis}~\cite{ravinderFinancialAnalysisStudy2013}.

\begin{itemize}
    \item
    \emph{Horizontal analysis} aims to analyze changes in different data points of the subject over time, with textual content containing timestamps and trend words. For example, \emph{`VC fundings have risen steadily over the past decade.'}
    Sometimes, there are only trend words but not timestamps, requiring analysis of the original data table, for example, \emph{`Dow Jones Index (US30) rises strongly.'}
    
    \item
    \emph{Vertical analysis} analyzes the relationship between different individual data points or between data points and their totals.
    \emph{Vertical analysis} usually involves numerical comparisons of multiple data points at the same timestamp, which is also referred to as static analysis~\cite{ravinderFinancialAnalysisStudy2013}. For instance, \emph{`China sold 2.1 million units of passenger new energy vehicles in the past quarter, with NEV penetrations reaching 31\%'.}
    
    \item
    \emph{Combined analysis} considers interrelated data points simultaneously, which is useful for complex situations in financial narratives, such as to compare multiple data items. 
    \emph{Combined analysis} is challenging, as it involves both the accurate identification of multiple data items within the text and their binding with values in the data table.
    It also requires the selection of appropriate combinations of graphical overlays to convey complex relationships.
    For instance, in the following narrative, \emph{`BYD and LG Energy Solution together hold about a quarter of the global EV battery market share'}, we first need to identify the two entities: BYD and LG Energy Solution in the data, and then consider the proportion of their combination in the market share.
\end{itemize}
Through the examination of typical examples within these analysis categories, we identified that the narrative structure of a financial narrative can usually be discerned by extracting three types of vocabulary in the textual content: \texttt{subject}, \texttt{trend}, and \texttt{numerical}.
\texttt{Subject} vocabulary represents the focal entities readers are expected to identify and use to associate textual descriptions and the data.
\texttt{Trend} vocabulary can be further categorized into three types: describing the change pattern, describing the special event, and describing the summary indicator. 
\texttt{Numerical} vocabulary refers to quantitative data that provide concrete numerical information to support narratives.
The three types of vocabulary serve as central points of interest in financial narratives, with \texttt{trend} and \texttt{numerical} information often used to elucidate and characterize when reading.

To verify the correctness of the classification, we presented some narrative examples with annotated vocabularies to some independent experts.
For example, for the narrative `VC fundings have risen steadily over the past decade', the subject \emph{`\texttt{VC fundings}'} and trend describing the change pattern \emph{`\texttt{risen steadily}'} are annotated.
The experts gave positive feedback, affirming that the inclusion of these words aids in expediting the comprehension of financial narratives.
At the same time, experts pointed out that among these three vocabulary types, \texttt{trend} is the most difficult to be identified in reading, because identifying \texttt{trend} requires some domain knowledge in \emph{horizontal analysis}.
In contrast, \texttt{subject} and \texttt{numerical} are usually corresponding to table columns, which can be feasibly addressed by LLMs.

\begin{figure*}
    \centering
    \includegraphics[width=0.99\textwidth]{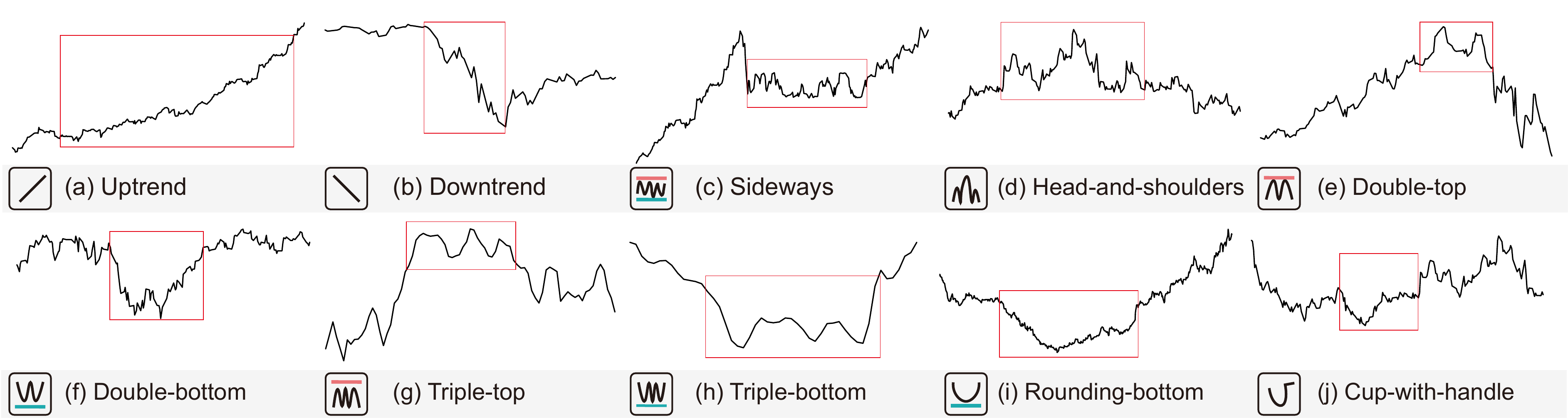}
    \caption{The top ten trend vocabularies summarized in the collected financial narrative dataset with their visual patterns.}
    \vspace{-2mm}
    \label{fig:pattern}
\end{figure*}

Next, to identify common \texttt{trend} vocabularies and enable LLMs to learn corresponding data patterns, we filtered financial narratives from the collected corpus with corresponding data tables available.
This yielded about 300 narratives from financial news and reports, whilst academic publications do not provide data table.
We further supplemented financial analyst reports from \emph{PitchBook}\footnote{\url{https://my.pitchbook.com/dashboard}}, which contain rich data tables with corresponding textual descriptions.
In the end, we collected 493 financial narratives with corresponding data tables.
Based on the collection, we summarized \texttt{trend} vocabularies from narratives, with the top 10 \texttt{trend} vocabularies and related data patterns shown in Figure~\ref{fig:pattern}.

\subsection{Correspondence between Graphical Overlays and Financial Narrative}
\label{ssec:correspondence}

\begin{figure}[t]
    \centering
    \includegraphics[width=0.99\linewidth]{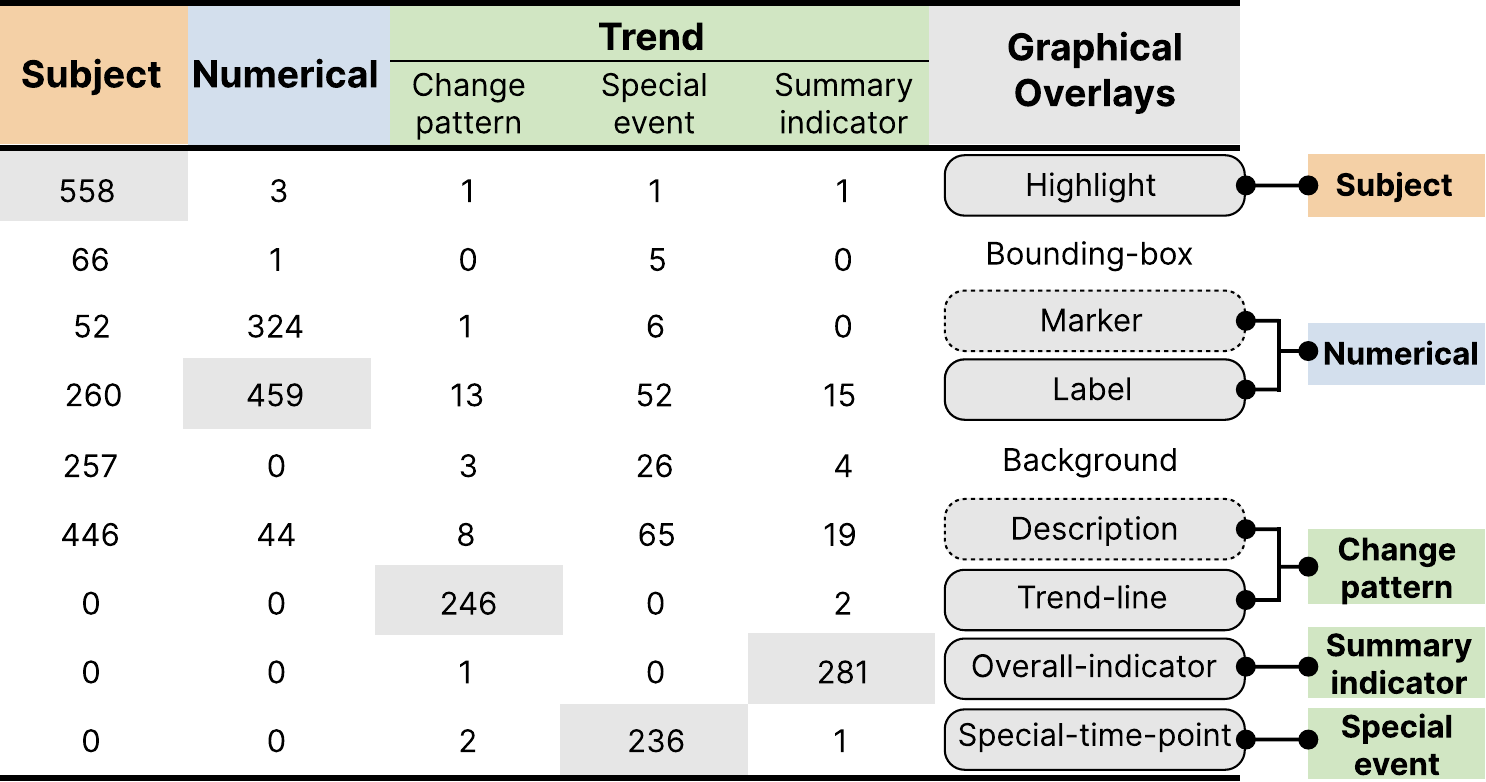}
    \vspace{-3mm}
    \caption{Statistics results on the correspondence between financial vocabularies and graphical overlays. The right shows the correspondence based on both the statistics and discussions with practitioners.}
    \vspace{-4mm}
    \label{fig:map}
\end{figure}

To determine the most suitable correspondence between graphical overlays and financial narratives, we conducted a two-phase approach of statistical analysis and expert interviews. We first performed a statistical analysis to examine the relationship between vocabulary and graphical overlay techniques in the collected corpus. The statistical results are shown on the left side of Figure~\ref{fig:map}. Then, we conducted interviews with four independent financial practitioners (3 males, 1 female, aged 22-34), each possessing over 4 years of experience in creating charts for financial analysis.
Considering both the statistical analysis and interviews, we derived the correspondence as illustrated in Figure~\ref{fig:map}:
\begin{itemize}
    \item The generated layered chart can incorporate a combination of graphical overlays.
    Statistical results indicate that most layered charts contain two or more graphical overlays. The practitioners also agreed that different graphical overlays can be employed for various financial vocabularies.

    \item The default graphical overlay for \texttt{subject} vocabulary is set to \emph{highlight}, the most used graphical overlay for \texttt{subject} vocabulary as shown in Figure~\ref{fig:map}.
    \emph{bounding-box} and \emph{background} are also very common but not selected, because the finance practitioners noted that compared with these two techniques, \emph{highlight} is better suited to all chart types.

    \item The default graphical overlay for \texttt{numerical} vocabulary is set to the combination of \emph{marker} and \emph{label}, which are both commonly used for annotating \texttt{numerical} in the collected layered charts. The practitioners also agreed with the choice, as this combination does not conflict in visual presentation and would help swiftly locate data points.

    \item The default graphical overlay techniques for \emph{trend} vocabulary are set to the combination of \emph{trend-line} and \emph{description} when describing the change pattern, \emph{overall-indicator} when describing the summary indicator, and \emph{special-time-point} when describing the special event, respectively.
    The combination of \emph{trend-line} and \emph{description} has appeared in the collected layered chart, and practitioners agreed that this combination could help readers swiftly understand the change patterns.

\end{itemize}

With the derived correspondence, we are able to apply suitable graphical overlays for different financial narratives.
For instance, the narrative \emph{`\texttt{VC fundings} have \texttt{risen steadily} over the past decade'}, has the subject vocabulary \emph{`\texttt{VC fundings}'} and the trend vocabulary \emph{`\texttt{risen steadily}'} describing change patterns.
As such, its default layered chart includes the combination of \emph{highlight}, \emph{description} and \emph{trend-line}.

\section{FinFlier}
\begin{figure*}
    \centering
    \includegraphics[width=0.99\linewidth]{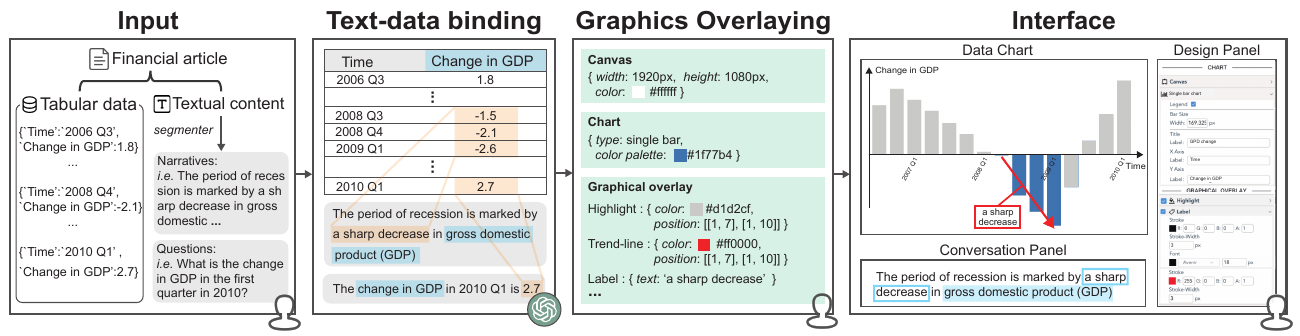}
    \vspace{-4mm}
    \caption{\tool system mainly consists of two main modules: \emph{text-data binding} module and \emph{graphics overlaying} module. \tool takes tabular data and textual content in a financial article as input. It first segments the textual content into narratives. For each narrative, \tool then extracts the template of text-data binding and passes the template to the {graphics overlaying} module, which outputs a layered chart with layering graphical overlays. Users can edit the generated layered chart, and continue questioning or feedback errors.}
    \label{fig:pipeline}
\end{figure*}

The workflow of \tool is depicted in Figure~\ref{fig:pipeline}, which contains two main modules and an interactive interface.
\tool takes a data table and textual content as inputs.
Through an LLM-based narrative segmenter, the textual content is segmented into narratives according to different \texttt{subjects}.
For each narrative, we leverage a knowledge-grounding LLM in the \emph{text-data binding} module (Sect.~\ref{ssec:connection}) to establish the connection between financial vocabulary and tabular data.
The \emph{graphics overlaying} module (Sect.~\ref{ssec:implementation}) generates effective layered charts, taking the derived correspondence between graphical overlays and financial narrative into account.
We further design an interactive interface to help users generate and iterate on layered charts with personalized design preferences (Sect.~\ref{ssec:interface}).

\subsection{Module 1: Text-data Binding}\label{ssec:connection}
\begin{figure*}
    \centering
    \includegraphics[width=0.88\linewidth]{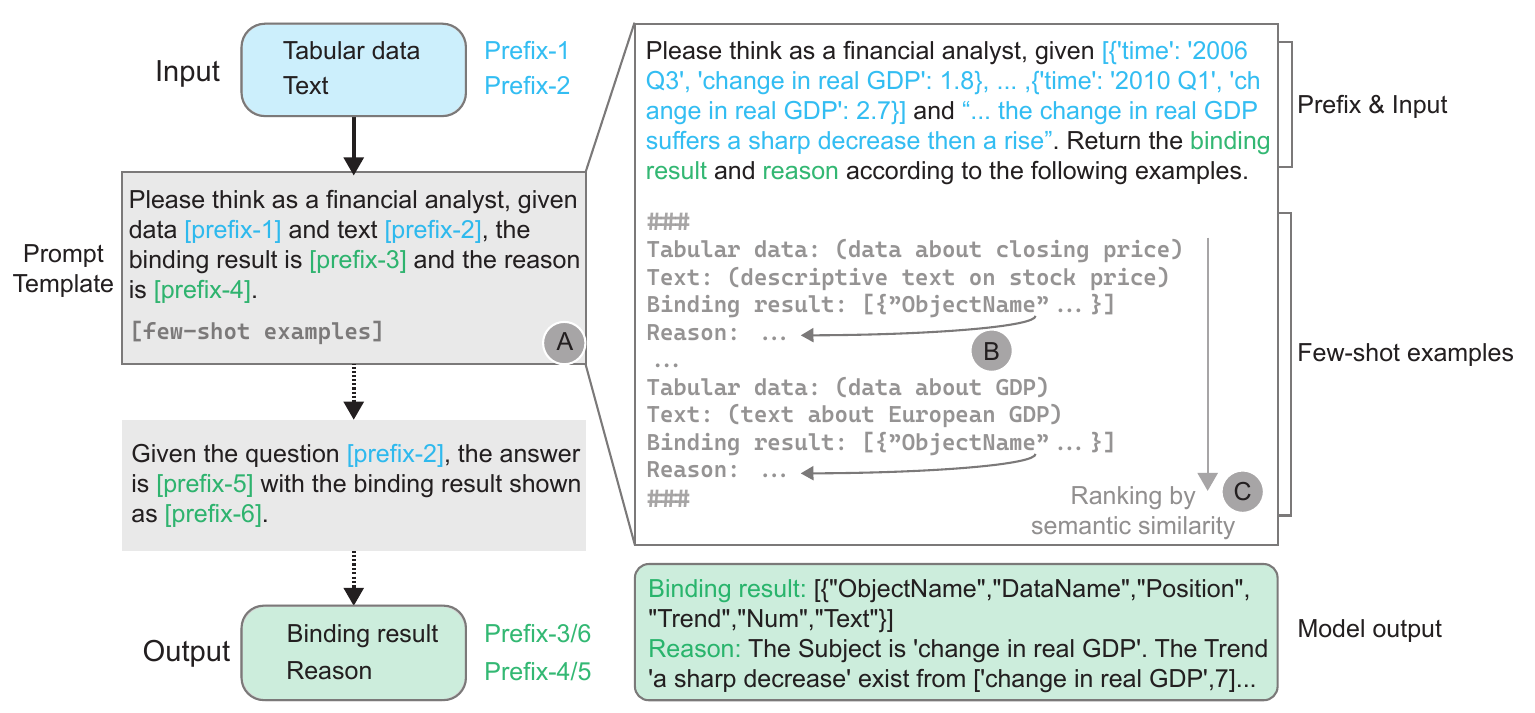}
    \caption{An example of the pipeline for \emph{text-data binding} module.
    The left part shows the pipeline of \emph{text-data binding} module, where three approaches are used to optimize the prompts, including (A) output constraint, (B) chain-of-thought, and (C) dynamic prompt. The template includes placeholders for the input tabular data (prefix-1), input text (prefix-2), output binding result (prefix-3/6), output reason (prefix-4/5), and (optional) few-shot examples.
    The right part shows the actual prompt after filling in the placeholders in the prompt template.}
    \label{fig:module1}
\end{figure*}

To address the complexity of narrative structures and vocabulary in financial narratives (Sect.~\ref{ssec:fin_pattern}), we incorporate financial domain knowledge into the base GPT-3.5 model through a series of prompt engineering techniques, including:

\noindent
\textbf{Output Constraint.} 
LLMs are sensitive to input prompts, tending to favor certain prompt formats~\cite{aichain_2022} and paraphrases~\cite{aiprom_2021}.
In particular, GPT often generates outputs with varied structures, which are unsuitable for binding with the data and visualization. It is essential to constrain its outputs to follow specific formats.
Output constraint emerges as a potent technique to exert fine-grained control over generated responses, aligning them more precisely with the guidelines and formats~\cite{structureoutput_2024}.
Based on the derived design space, we create a collection of templates that encapsulate the output format.
By employing these templates, we transform the text-data binding task into a masked language modeling problem, wherein specific portions of the template are marked as masks to be filled by the model.
This approach enforces a predefined structure on the generated content, aligning it with our intended output format.
The structured template can be seen in Figure~\ref{fig:module1} (A), where the input tabular data pertains to the change in real GDP, while the input text describes the change pattern in GDP, showing a decline followed by a rise during the recession period.
The template includes six objects of `ObjectName', `DataName', `Position', `Trend', `Num', and `Text' as outputs by the knowledge-grounding LLM.
The `ObjectName' corresponds to the \texttt{subject} in the input text, while `DataName' represents the column in the data table corresponding to \texttt{subject} in the given tabular data.
`Trend' and `Num' correspond to the \texttt{trend}, and \texttt{numerical} extracted from the input text, respectively.
These two fields are mutually exclusive and do not have values simultaneously.
`Position' contains a variable-length array that represents the starting and ending positions of \texttt{trend}, or the positions of \texttt{numerical} in the given data table.
The utilization of templates not only constrains the generated content but also guides the model's attention toward the relevant components.
This increased focus enhances the accuracy and appropriateness of generated responses.
The corresponding binding result for the template and more result examples can be viewed in supplementary A.

\noindent
\textbf{Chain-of-Thought (CoT).}
LLMs may struggle on tasks like recognizing and tasks requiring branching logic~\cite{logic_2021,logic_2020}, because they are designed to grasp language form, rather than the meaning~\cite{weakllm_2020}.
In the domain of prompt optimization, the ``chain-of-thought" technique emerges as a transformative strategy to dissect complex reasoning tasks, bolster the model's comprehension, and elevate its response generation capabilities~\cite{cot_2022}.
Our text-data binding task can be viewed as a reasoning task where the model needs to populate the template based on the given text and data.
So, we introduce CoT for a more accurate reasoning process (Figure~\ref{fig:module1} (B)).
We collaborate with finance domain experts to identify the reasoning processes for populating templates, especially for special change patterns.
These reasoning steps are subsequently integrated into the prompt sequence, guiding the model's generation process along a structured path.
The detailed reasoning content filled in an example can be viewed in supplementary A.
Adopting the CoT technique not only improves the model's reasoning ability to achieve better text-data binding accuracy, but also helps users gain insights into the underlying logic and point out errors in the reasoning process.
This interactive feedback loop fosters a more effective interaction with the model.

\noindent
\textbf{Dynamic Prompt.}
Traditional few-shot prompting can lead to poor response quality when there is a large divergence between the provided case and the prompt examples.
Furthermore, the order of providing examples within prompts has been proven to significantly impact the performance of LLMs~\cite{prompteng_2023}.
Building on insights from the evaluation results of LM-BFF~\cite{LM-BFF_2020}, we recognize that the detrimental effects of fixed-order can be mitigated by dynamically constructing prompts based on semantic similarity to bridge the gap between examples and input cases and result in correct responses.
To implement this concept, we only sample demonstrations that are closely related to the input.
After the user enters the text, we retrieve the top-k most relevant prompt examples from the constructed prompt database and sort them to create a customized prompt sequence that better matches the user's input. In our implementation, we set k as 10.
As Figure~\ref{fig:module1} (C) shows, for the input `GDP change' case, \tool puts an example about European GDP at the end of the few-shot examples.

\begin{figure*}[t]
    \centering
    \includegraphics[width=0.995\textwidth]{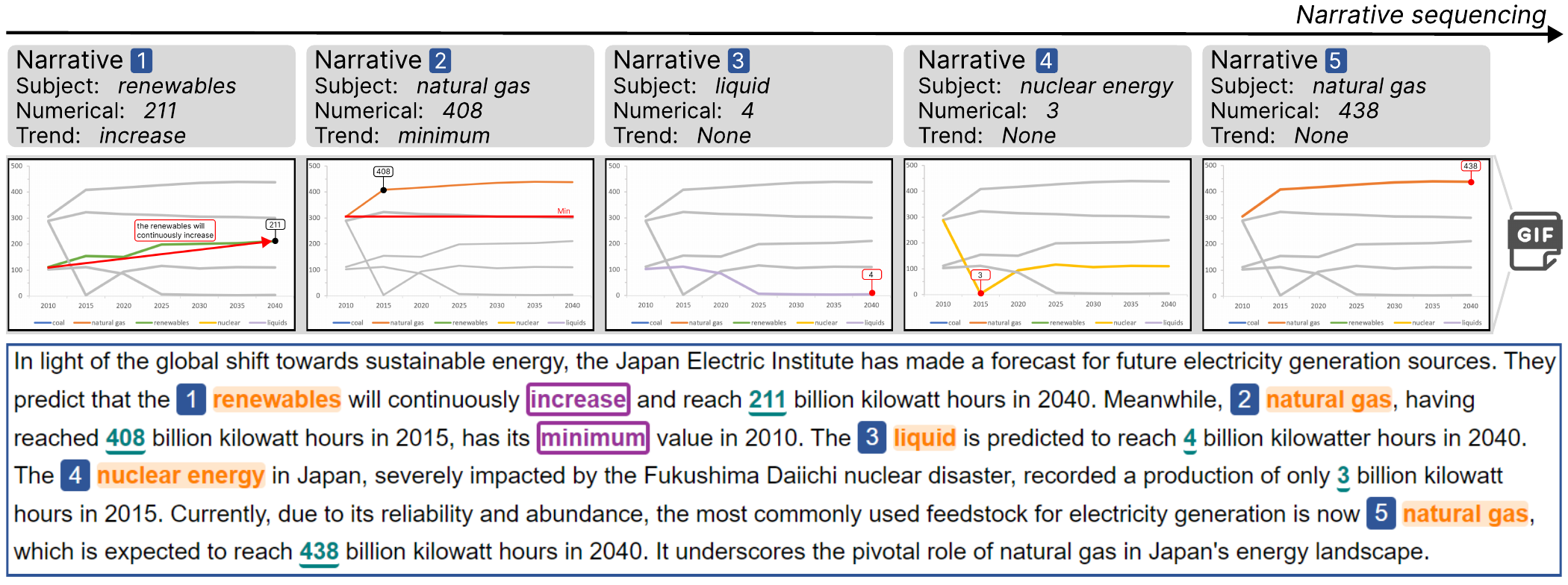}
    \vspace{-3mm}
    \caption{The output of \tool conforms to the narrative format. Five automatically generated layered charts present a complete narrative of Japan Electric Institute's forecasts for power generation from different sources. \tool supports exporting individual layered charts in PNG format, as well as exporting a sequence of all layered charts according to the narrative sequencing as a GIF file.}    
    \vspace{-3mm}
    \label{fig:output}
\end{figure*}

The synergy between output constraint, CoT, and dynamic prompt presents a comprehensive solution to the challenges posed by the inherent flexibility of large language models. 
By integrating these techniques, we craft a financial knowledge-grounding LLM for text-data binding, which allows \tool to provide users with accurate and contextually relevant binding results.

\subsection{Module 2: Graphics Overlaying}\label{ssec:implementation}
The primary component of \emph{graphics overlaying} is layering graphical overlays, where we use a canvas element placed over the base chart.
This section describes the implementation details including the default position (Sect.~\ref{sec:position}) and color (Sect.~\ref{sec:color}) for each type of graphical overlays.
In the end, we describe how to sequence the generated layered charts to guide users through a whole financial article (Sect.~\ref{sec:sequence}).

\subsubsection{Positioning}\label{sec:position}
Effective positioning of graphical overlays is crucial in enhancing the clarity and interpretability of generated charts.
\tool provides default position based on heuristics and previous work~\cite{graphicalo_2012}, as overviewed in Figure~\ref{fig:go_map}.
When adding \emph{marker}, \emph{label}, or \emph{description} to charts, the default positions are strategically specified relative to the chart boundaries and allow subsequent user adjustments.
The default \emph{marker} is a circle with a radius of 2 pixels and is placed at the top center of the bar or the data point on the line.
The default \emph{label/description}, formatted within a rectangle boundary, aligns with the top edge of the chart and is positioned perpendicular to the \emph{marker}.
For \emph{overall-indicator}, we introduce a parallel line plotted on the chart according to its statistical value (\eg, mean, global maximum, and global minimum).
The corresponding statistic label is placed above the line and centered horizontally within the chart area, providing a quick and clear reference point for the readers.

\subsubsection{Color}\label{sec:color}
The judicious use of color in graphical overlays is pivotal in ensuring that the visual elements intended to be highlighted are indeed prominent and easily discernible.
\tool provides a default color palette that aligns with user preferences and enhances the visual impact of the charts.
For the color palette of graphical overlays, two prevalent methods are commonly employed: highlighting elements by desaturating non-highlighted ones, and using highly saturated contrasting colors to break the Gestalt principle of similarity~\cite{gestalt_1938}.
Each method has its merits and is suited to different graphical techniques.
We apply desaturation to \emph{color} and \emph{background} techniques as it prevents the visual confusion that can arise from an overabundance of colors in the chart.
On the other hand, the use of highly saturated contrasting colors, including red and black, is particularly effective for graphical overlay techniques like \emph{bounding-box}, \emph{marker}, etc.
These colors create a clear visual break, making the highlighted elements instantly recognizable.
The default colors for each graphical overlay technique can be overviewed in Figure~\ref{fig:go_map}.

\subsubsection{Chart Sequencing}\label{sec:sequence}
The \textit{text-data binding} module focuses on the recognition of three distinct vocabularies within the textual content and subsequently establishes a binding between data and these vocabularies.
A narrative, as we defined, is a segment of textual content that corresponds to a single \texttt{subject}.
Each financial article inherently possesses an implicit narrative sequencing determined by the sequence where different \texttt{subjects} are introduced.
This sequencing is crucial as it represents the order in which users' attention shifts in visualization.
Within each narrative, the arrangement of \texttt{trend} and \texttt{numerical} values influence the overlays of the generated layered chart.
As illustrated in Figure~\ref{fig:output}, a given financial article contains five narratives. The first narrative ``\emph{They predict that the renewables will continuously increase and reach 211 billion kilowatt hours in 2040.}'' describes the subject \texttt{renewables} with the trend \texttt{increase} and numerical \texttt{211}. The corresponding generated layered chart helps users swiftly focus on this narrative through layering graphical overlays.

To reflect the narrative sequencing, \tool enables to generate multiple layered charts ordered by the narrative sequence.
As illustrated in Figure~\ref{fig:output}, the financial article comprises sequential narratives, with each narrative enclosed within its own layered chart. 
The layered charts are sequenced together, providing users with comprehensive and logically structured financial visualizations.
The array of layered charts can be effectively organized into a GIF output, facilitating a comprehensive grasp of the entire financial article.
In this way, chart sequencing improves users' understanding of the financial narrative details, meanwhile offering a rapid overview of the entire article.

\begin{figure*}
    \centering
    \includegraphics[width=0.99\textwidth]{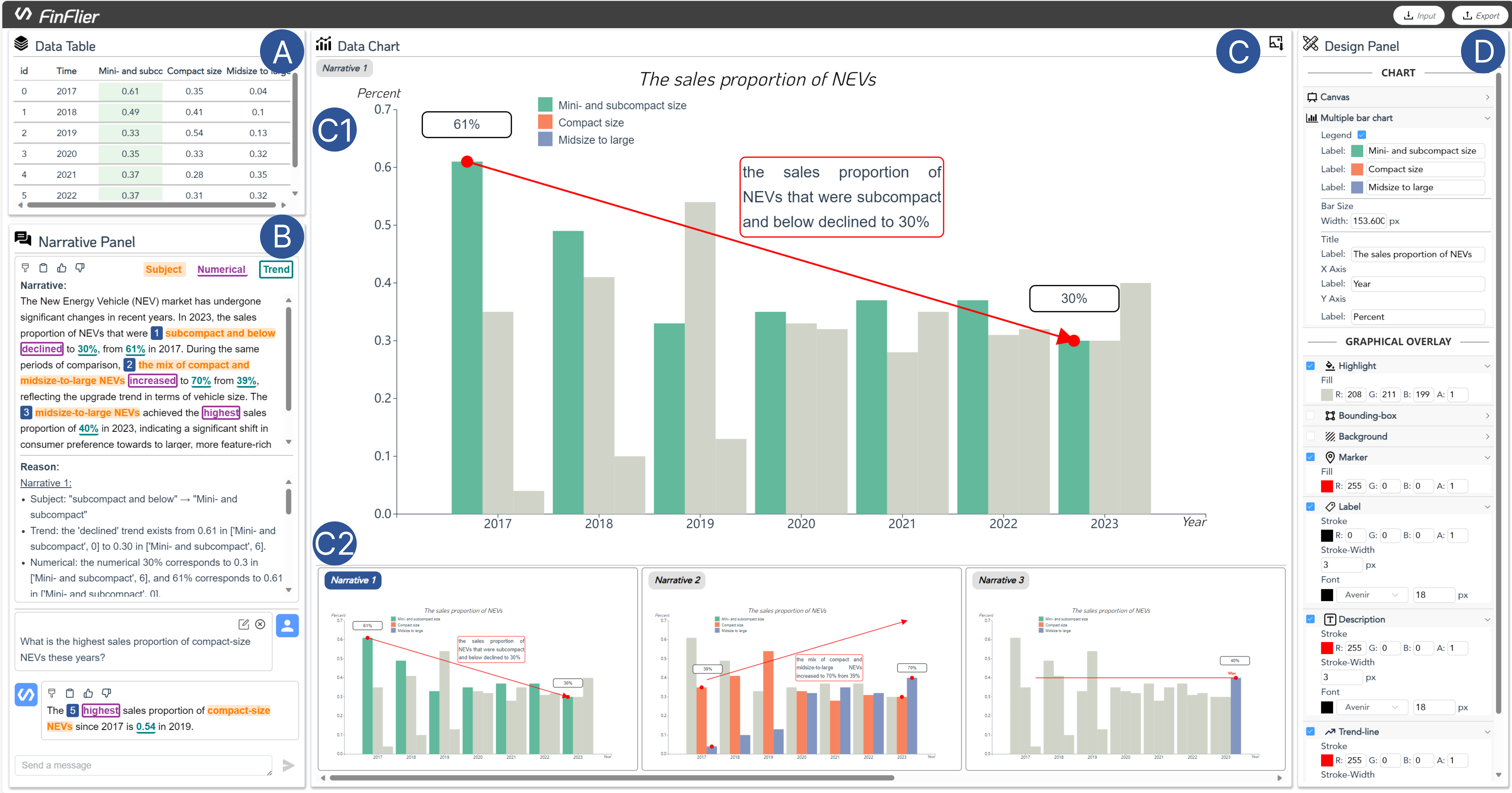}
    \vspace{-3mm}
    \caption{The \tool user interface includes (A) Data Table that connects data points to the original tabular data, (B) Narrative Panel for natural language inputs and responses, (C) Data Chart that provides the generated layered chart and narrative overview, and (D) Design Panel that provides operations for canvas, visual elements and graphical overlays.}
    \vspace{-3mm}
    \label{fig:interface}
\end{figure*}

\subsection{Interactive Interface}\label{ssec:interface}
We design an interface for our system which consists of four views, as shown in Figure~\ref{fig:interface}: (A) Data Table, (B) Narrative Panel, (C) Data Chart, and (D) Design Panel.
When a user uploads a data file with its corresponding textual content, the interface presents these views in a coordinated manner.

In Data Table view (Figure~\ref{fig:interface} (A)), the corresponding data table is displayed.
Narrative Panel (Figure~\ref{fig:interface} (B)) displays the annotated textual content, where the background color of \texttt{subjects} is added, \texttt{\uline{numerical}} values are underlined, and \texttt{\boxed{trend}} words are boxed.
Additionally, \tool provides the reasoning behind the recognition in Narrative Panel to assist users in understanding the basis for the text-data bindings.
Data Chart view (Figure~\ref{fig:interface} (C)) showcases the base chart. Upon selecting a narrative, \tool automatically generates and displays the corresponding layered chart (Figure~\ref{fig:interface} (C1)).
Below the generated layered chart is a chart gallery for all narratives (Figure~\ref{fig:interface} (C2)).
For further exploration, users can continue to pose questions about the uploaded article and they will receive responses in Narrative Panel, facilitating further analysis.
Design Panel (Figure~\ref{fig:interface} (D)) provides operations of canvas, chart elements, and graphical overlay techniques, enabling users to flexibly combine different techniques and customize visual elements.
These coordinated views enable users to quickly create and iterate financial narratives, and enhance their comprehension of the text-chart relationship.

\begin{figure*}
    \centering
    \includegraphics[width=0.99\textwidth]{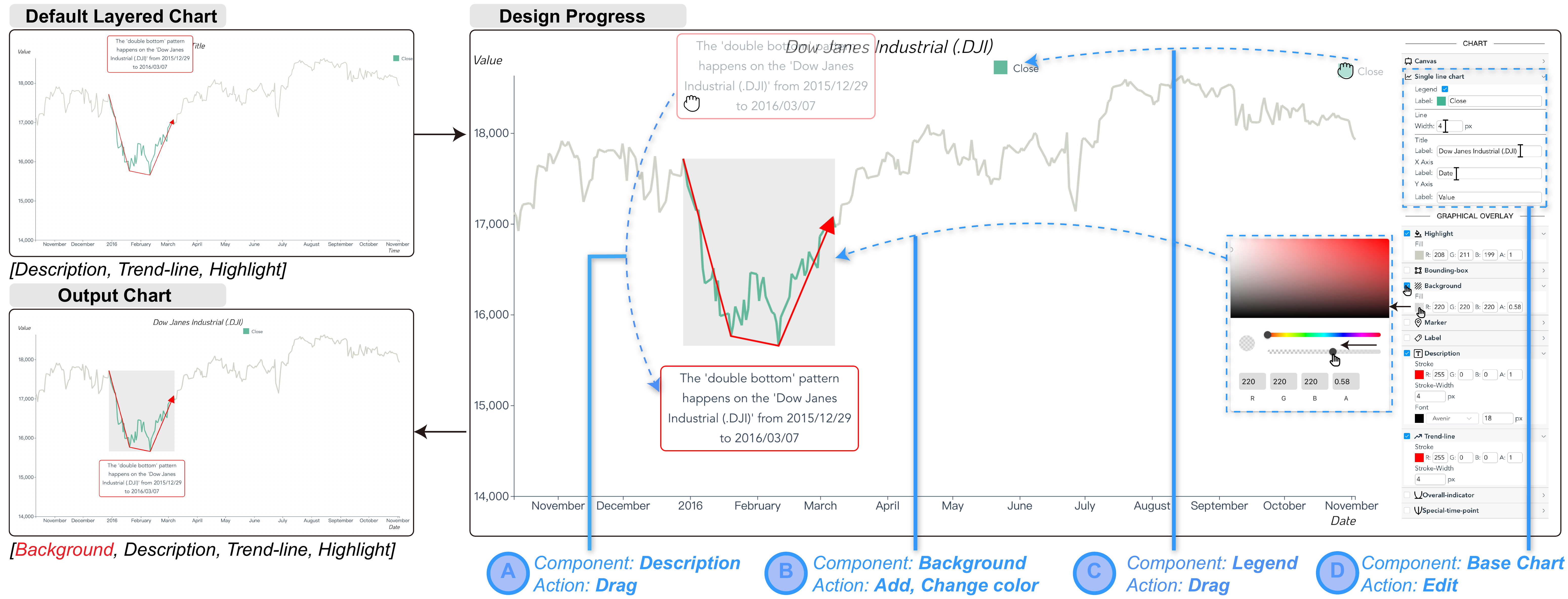}
    \vspace{-2mm}
    \caption{Based on the default layered chart, the user can flexibly edit it through \emph{Design Panel}. In this example, the user (A) \emph{Drag} the position of \emph{description}, (B) \emph{Add} the \emph{background} and \emph{Change color}, (C) \emph{Drag} the position of legend, and (D) \emph{Edit} the line width, chart title, and X axis name. Finally, the user gets a customized layered chart.}
    \label{fig:interaction}
\end{figure*}

\vspace{1.5mm}
\noindent
\textbf{Interactions for text-data binding}.
\tool interface enables users to engage through natural language in Narrative Panel.
After entering textual content,  \tool offers two primary operations:
\begin{itemize}
    \item Editing \img{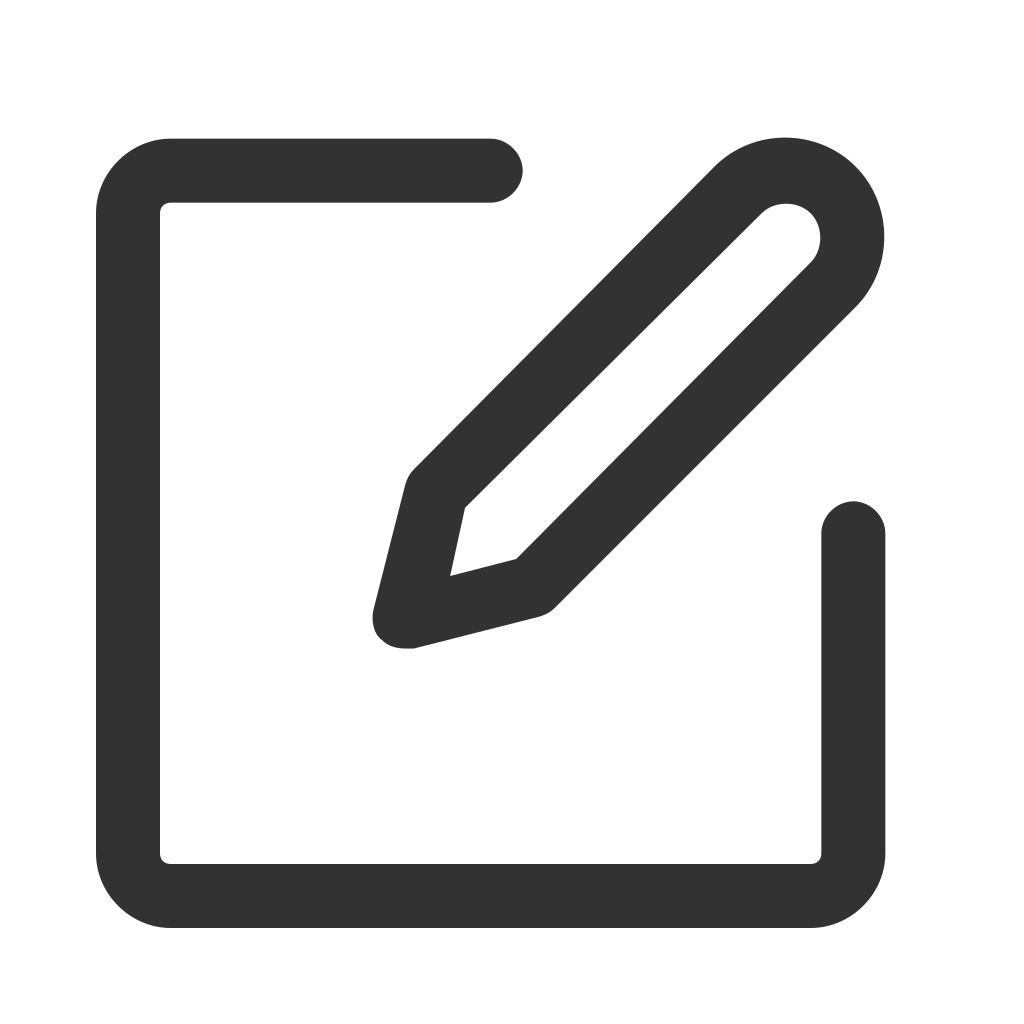}: When users click on the Editing button, they can revise the input textual content and subsequently send it to the system to obtain a new response.
    \item Deletion \img{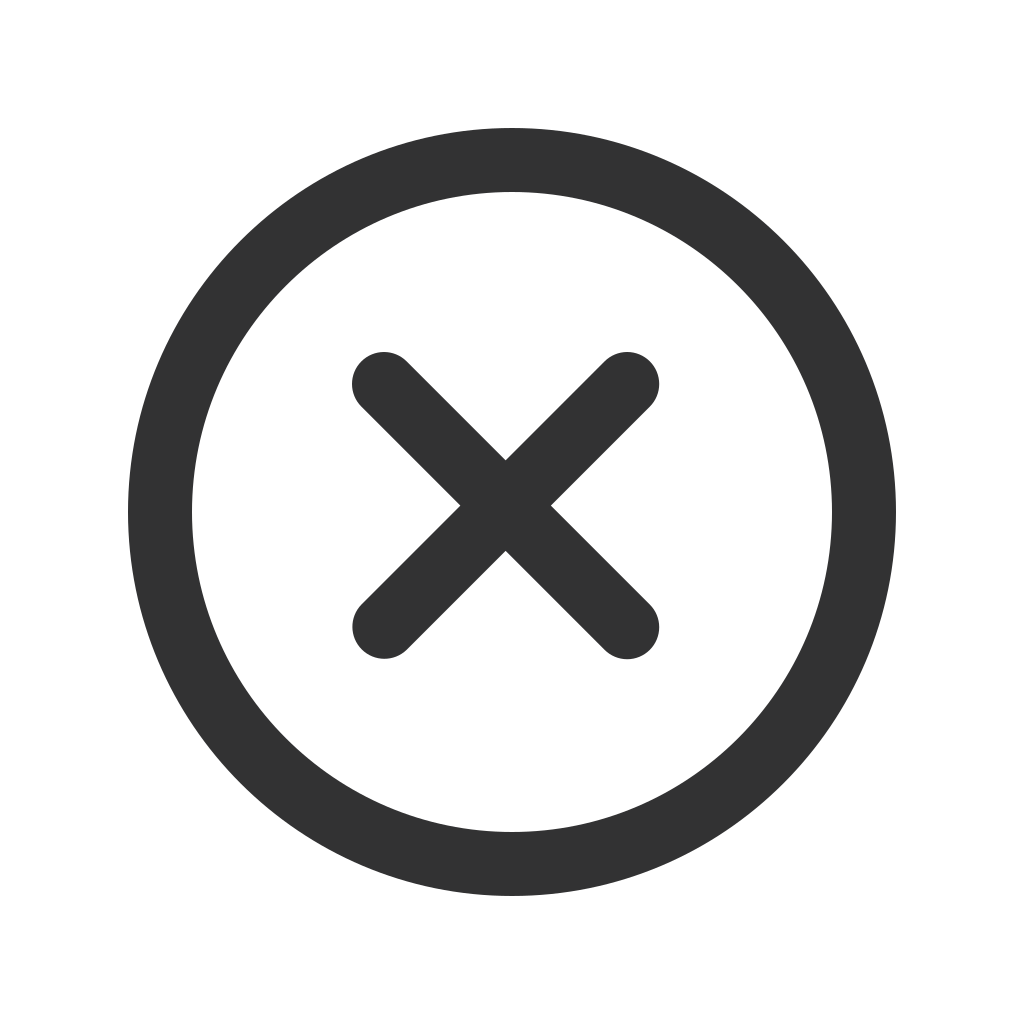}: Clicking on the Deletion button removes both the input content and its corresponding response. 
\end{itemize}

We provide four buttons for user feedback on the text-data binding results:
\begin{itemize}
    \item Mark \img{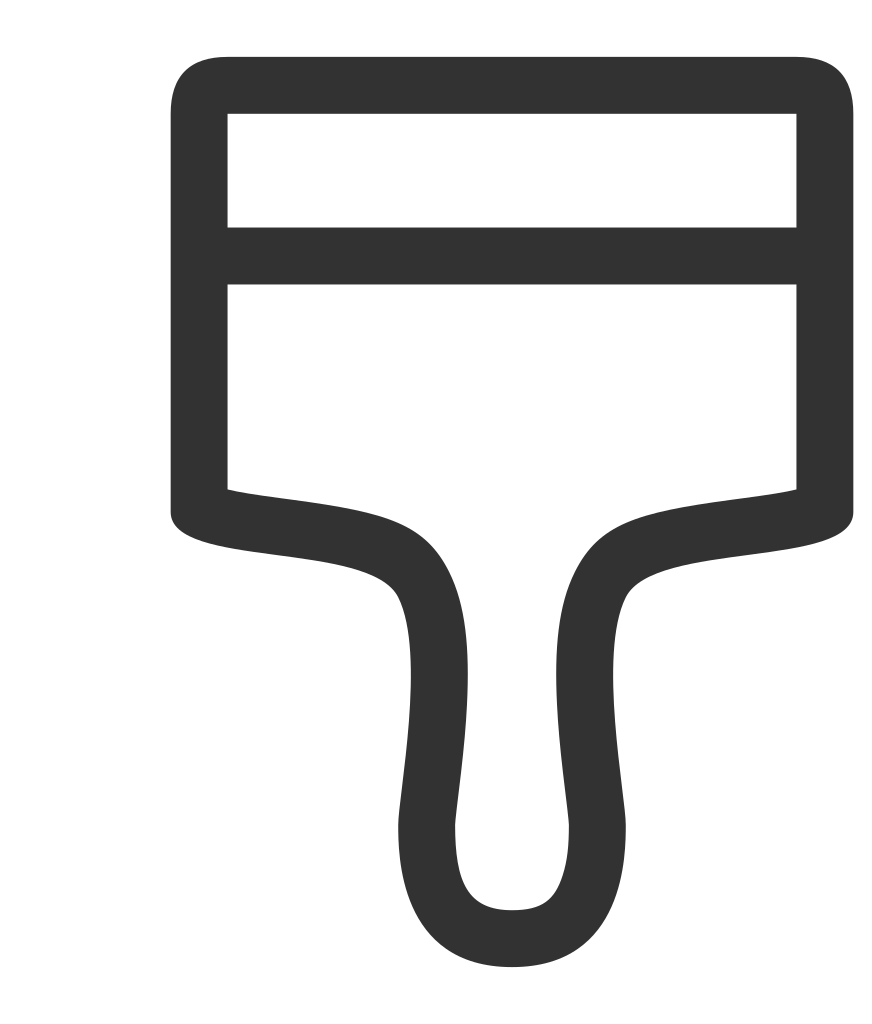}: When users choose the Mark button, they can select any parts of the response that they believe are incorrectly labeled or missed in the \emph{text-data binding} module. Subsequently, this text-data pair is stored for later expert review. We periodically curate the stored text-data pairs and put representative examples into our prompt example datasets, aiming to minimize similar errors in future interactions.
    \item Copy \img{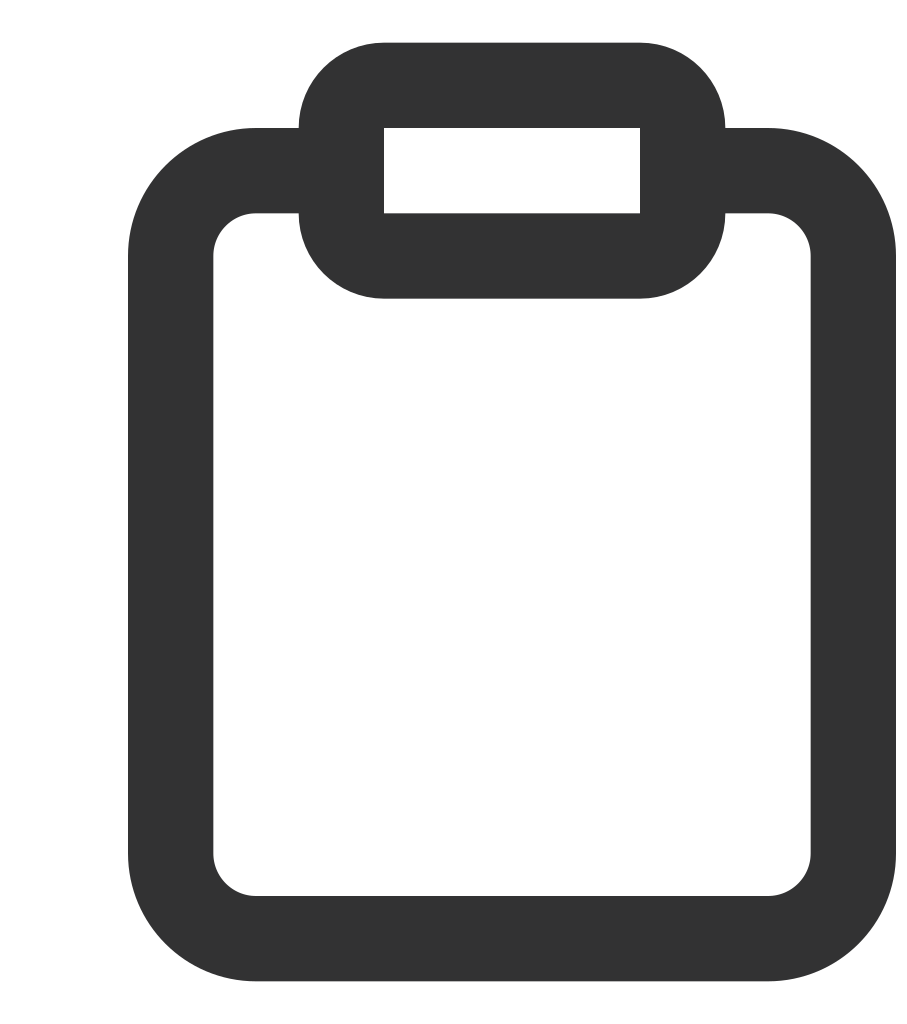}. Copy button enables users to copy the system's response to their clipboard for easy reference.
    \item Thumb-up \img{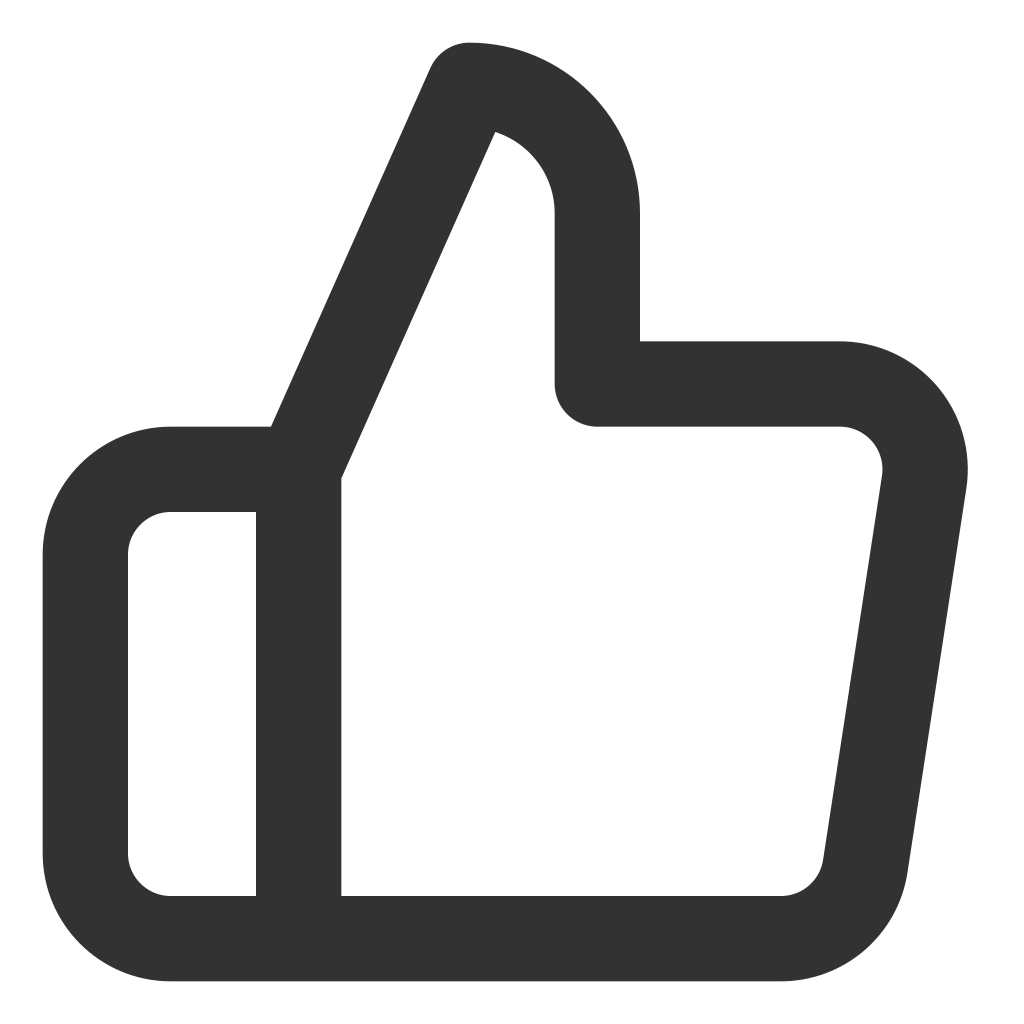}. If a user is satisfied with the response, s/he can express this sentiment by clicking the Thumb-up button.
    \item Thumb-down \img{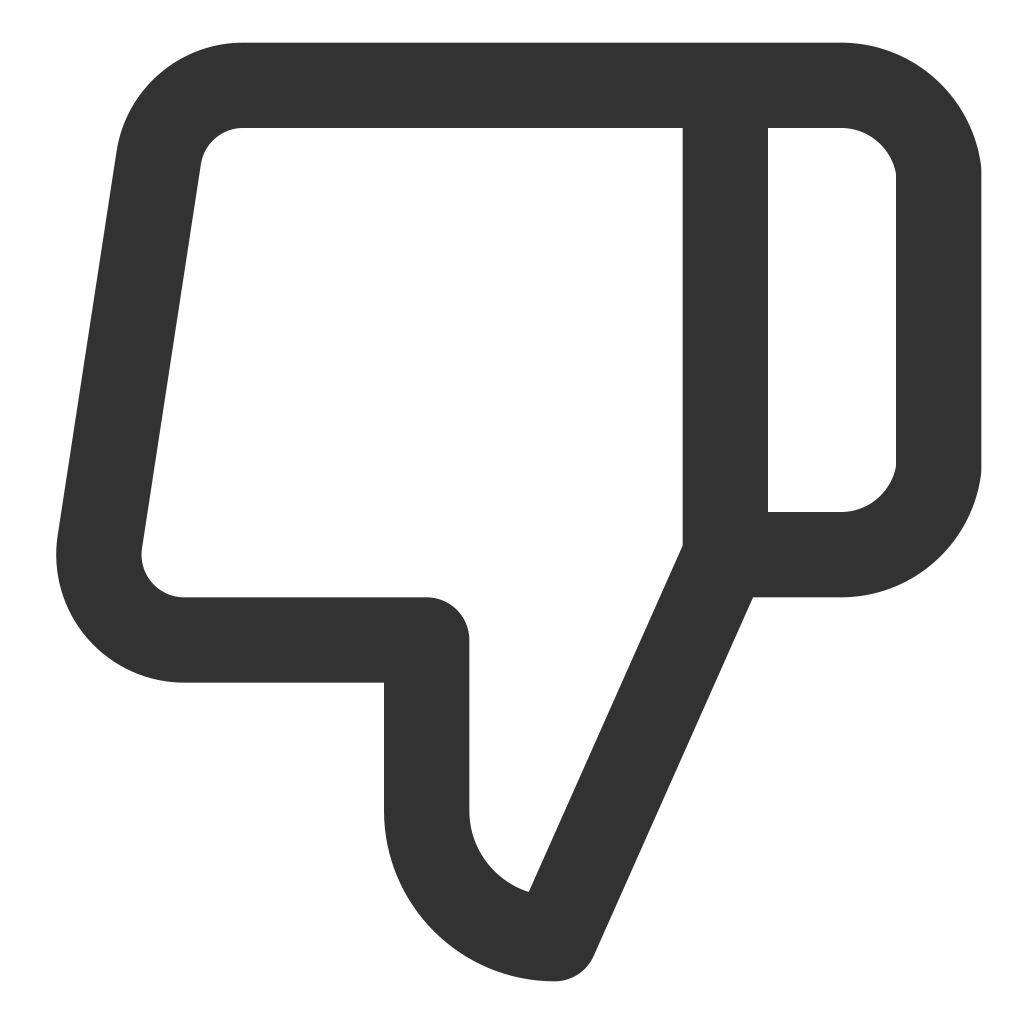}. If the response does not satisfy a user's expectations, s/he can click the Thumb-down button and receive a freshly generated response from the system.
\end{itemize}

\vspace{1.5mm}
\noindent
\textbf{Interactions for graphics overlaying}.
Once the default layered chart is generated, the canvas size, visual elements, and graphical overlays can be adjusted using \tool's Design Panel. 
Users can have full control over the canvas size, visual elements for base chart, and graphical overlays.
They can make flexible adjustments, including altering the color and width of bars, modifying the chart title, and editing the axis labels.
By selecting a combination of graphical overlay techniques and adjusting their properties, users can tailor the generated layered chart to align seamlessly with the currently selected narrative in Data Chart.
This flexibility ensures that the visual representation accurately conveys the intended narrative.
For those seeking a static image (in PNG format) of the generated layered chart, a convenient Save button 
located in the upper right corner of Data Chart (Figure~\ref{fig:interaction} (C)) is available.
With a single click, users can export the layered chart shown in Data Chart as a static PNG image.
Through clicking on the Export button 
located in the upper right corner of the interface, \tool allows for the export of static images for each individual narrative and all narratives as a GIF, preserving the narrative sequencing.

For instance, given a narrative about the Dow Janes Industrial (.DJI) undergoing a double-bottom pattern, \tool identifies the subject \texttt{Dow Janes Industrial (.DJI)} with the trend \texttt{double-bottom}, and automatically generates a default layered chart with a combination of \emph{highlight}, \emph{description}, and \emph{trend-line}.
However, users may have specific preferences for chart layout.
In the default chart, the position of description text is directly above the trend, which obscures the chart title, and the legend of chart is located in the top right corner of the chart.
To address this, users can easily drag the description text to a blank area (Figure~\ref{fig:interaction} (A)) and reposition the legend to the center of the chart (Figure~\ref{fig:interaction} (C)).
Users may also wish to highlight specific areas of the double-bottom pattern, which can be achieved by adding \emph{background} to the current combination of graphical overlays (Figure~\ref{fig:interaction} (B)). 
Additionally, they have the option to edit the visual elements within the base chart (Figure~\ref{fig:interaction} (D)) to better align with their specific requirements.
After customizing the satisfied layered chart, users can click the Save button to export it as a single PNG image.
\section{Evaluation}
To comprehensively evaluate the effectiveness of \tool, we illustrate examples by the system (Sect.~\ref{ssec:case}), conduct quantitative evaluation for text-data binding (Sect.~\ref{ssec:eva_txt-data}), and perform qualitative evaluation for the system (Sect.~\ref{ssec:user_study}).

\subsection{Examples}\label{ssec:case}
\begin{figure*}[htbp]
  \centering
  \includegraphics[width=0.99\linewidth]{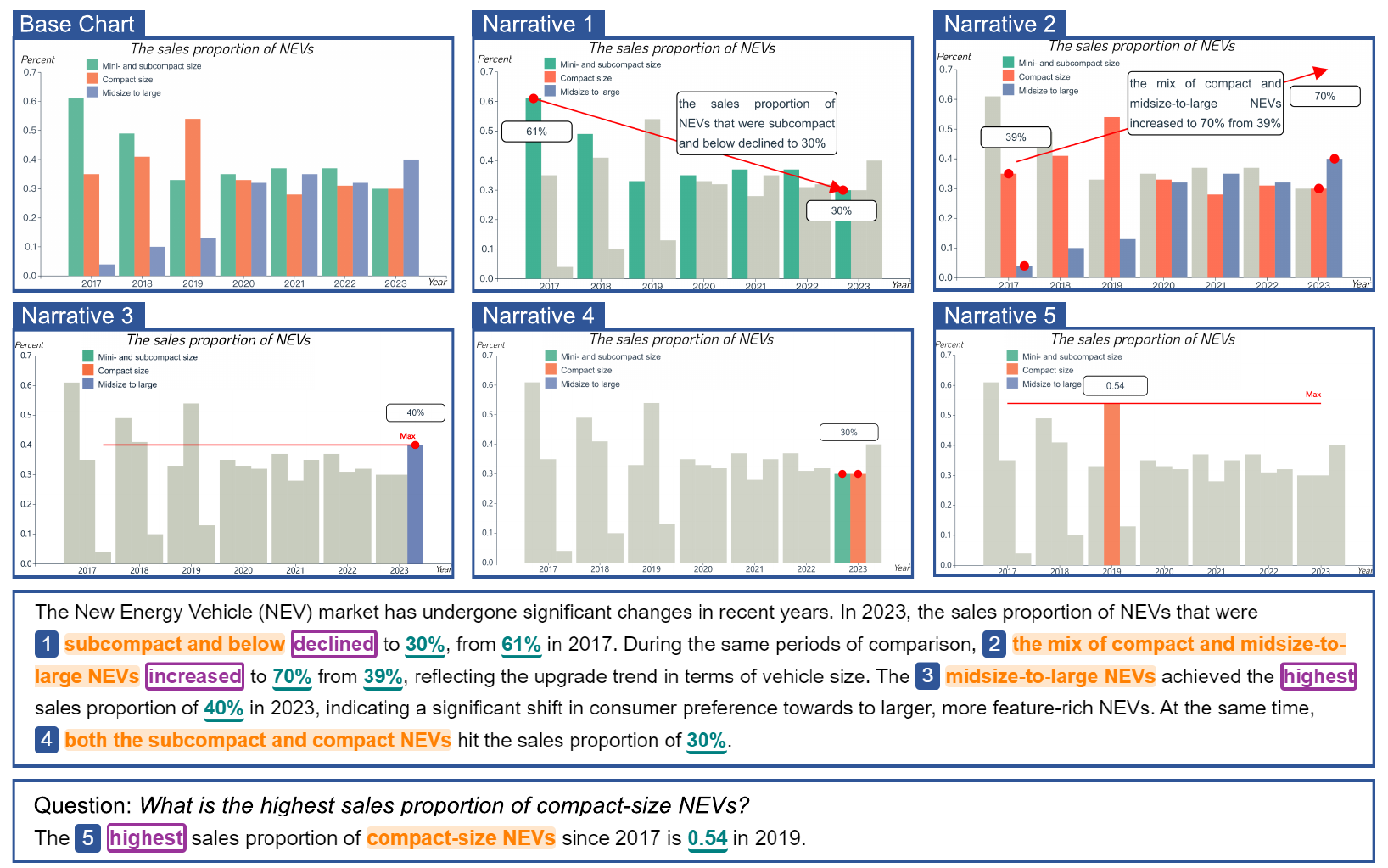}
  \vspace{-3mm}
  \caption{A sequence of automatically generated layered charts of the sales proportion of NEVs from 2017 to 2023.
  Each narrative comprises the \texttt{subject}, \texttt{numerical} and \texttt{trend} words, which are labeled in textual content and displayed in layered charts with default graphical overlays.
  }
  \label{fig:teaser}
\end{figure*}

To evaluate \tool's effectiveness and feasibility in the finance domain, we illustrate examples within these three analysis scenarios, \ie, horizontal, vertical, and combined analysis.
We select a financial narrative which details the changes in the new energy vehicle market over these years, encompassing all three analysis scenarios.
The narrative and corresponding layered charts generated by \tool are presented in Figure~\ref{fig:teaser}, with the base chart shown in the top-left corner.

\vspace{1mm}
\noindent
\emph{Horizontal analysis}.
Narrative 1 presents a typical \emph{horizontal analysis} scenario, reflecting the trend \texttt{declined} for the subject `\texttt{subcompact and below}' from 2017 to 2023. 
It also mentions numerical values \texttt{30\%} and \texttt{61\%}.
As such, the default combination of graphical overlays includes \emph{highlight}, \emph{trend-line}, \emph{description}, \emph{label} and \emph{marker}.

\vspace{1mm}
\noindent
\emph{Vertical analysis}.
Narrative 3 and Narrative 5 depict \emph{vertical analysis} scenarios.
They focus on showcasing specific values \texttt{40\%} for subject `\texttt{midsize-to-large NEVs}' and \texttt{0.54} for the subject `\texttt{compact-size NEVs}'.
For these two narratives, the default combination of graphical overlays includes \emph{highlight}, \emph{marker}, \emph{label}, and \emph{overall-indicator}.
The generated layered charts help users quickly locate specific data points, facilitating a detailed vertical analysis.

\vspace{1mm}
\noindent
\emph{Combined analysis}.
Narrative 2 and Narrative 4 introduce the combined analysis scenarios.
In Narrative 2, the subject \texttt{the mix of compact and midsize-to-large} combines data items from two different columns: `Compact size' and `Midsize to large'. 
This combined subject corresponds to a value of \texttt{70\%} in 2023, representing the total percentage of the two data values in this year.
Without the assistance of the automatically generated charts, readers would need to decipher the combined trend by understanding that it is related to the summation of the values in both columns for year 2023.
They would then have to compare the heights of bars to mine \texttt{increased} trend.
Besides, in Narrative 4, the subject \texttt{both the subcompact and compact NEVs} combines data items from columns `Subcompact and below' and `Compact size'. These two columns all hit \texttt{30\%} in 2023. 
With the layered chart, readers can swiftly access all the relevant information, enabling them to grasp the nuances of the combined subject in a short time.

\subsection{Quantitative Evaluation for Text-data Binding}\label{ssec:eva_txt-data}
To evaluate the \emph{text-data binding} module, we conducted a comparative analysis of our knowledge-grounding GPT-3.5 against two base LLMs: zero-shot GPT-3.5 and zero-shot GPT-4.
We randomly selected 50 financial narratives with labeled vocabularies.
Additionally, we performed ablation experiments to assess the impact of the CoT and dynamic prompt techniques on our model, denoted as `Knowledge-grounding GPT-3.5 w/o CoT' and `Knowledge-grounding GPT-3.5 w/o DP', respectively.
Since the output constraint serves as the template rather than influencing the model's reasoning process, we focused on CoT and dynamic prompt techniques.

\noindent
\textbf{Measures}.
We employed three evaluation metrics - Precision (P), Recall (R) and F1-score (F1).
They are calculated as:
$P=TP/(TP+FP)$, 
$R=TP/(TP+FN)$,
and $F1 = 2 \cdot (P \cdot R) / (P+R)$,
where $TP$ represents true positives, indicating the number of correctly identified vocabularies by \emph{text-data binding} module; $FP$ represents false positives, denoting the number of non-vocabularies identified as vocabularies; and $FN$ stands for false negatives, indicating the number of vocabularies present in text but not identified by module.

\vspace{1mm}
\noindent
\textbf{Results}. 
As shown in Table \ref{tab:eva}, our method archives the highest F1 scores across all three vocabulary types, demonstrating the effectiveness of the knowledge-grounding LLM in generating credible text-data binding results.
Notably, compared to the two base LLMs, the improvement by knowledge grounding is particularly evident in the recognizing \texttt{trend} words and \texttt{numerical} vocabularies.
Interestingly, the GPT-4 model achieves the lowest precision in recognizing \texttt{numerical} values. 
Through investigation, we found that GPT-4 interprets its reasoning by setting trend-related values as \texttt{numerical}, which is unsuitable for text-data binding tasks.
This highlights the importance of integrating domain-specific knowledge into LLMs to enhance their ability to accurately interpret and bind textual content to relevant data.

For ablation experiments, the performance of Knowledge-grounding GPT-3.5 w/o DP appears to perform worse than base LLMs, suggesting that simply providing various examples during prompt engineering does not necessarily improve result accuracy.
This comparison underscores the potential of using dynamic prompt techniques to enable LLMs to learn from more relevant examples and produce higher-quality results.
Knowledge-grounding GPT-3.5 w/o CoT also underperforms our model, and the lack of reasoning may reduce user confidence in our results.

Nevertheless, our \emph{text-data binding} module also faced some failure cases as LLM-based methods encounter hallucination - generating incorrect or unfounded information~\cite{hall_survey}.

\begin{table}[t]
\centering
\renewcommand{\arraystretch}{0.99}
\caption{Performance comparison between our knowledge-grounding model with other models on financial vocabulary identification}
\label{tab:eva}
\begin{tabular}{@{}ccccc@{}}
\toprule
                                         &    & \texttt{Subject} & \texttt{Trend}  & \texttt{Numerical} \\ \hline
\multirow{3}{*}{\shortstack[c]{Zero-shot\\ GPT-3.5}}                 & P  & 0.8921       & 0.9130      & \textbf{1.0000}         \\
                                         & R  & 0.9615       & 0.8077      & 0.6364         \\
                                         & F1 & 0.9091       & 0.8571      & 0.7778     \\ \hline  
\multirow{3}{*}{\shortstack[c]{Zero-shot\\ GPT-4}}                 & P  & \textbf{0.9259 }      &   0.7667    &    0.9048      \\
                                         & R  &  0.9615      & \textbf{0.8846}      & \textbf{0.8636}         \\
                                         & F1 & 0.9434       & 0.8214      & 0.8837     \\ \hline       
\multirow{3}{*}{\shortstack[c]{Knowledge-grounding\\ GPT-3.5 w/o CoT}} & P  &  0.8929     &   0.8846    &   0.9000       \\
                                         & R  &  0.9615      & \textbf{0.8846}      & 0.8182         \\
                                         & F1 & 0.9259       & 0.8846      & 0.8571     \\  \hline    
\multirow{3}{*}{\shortstack[c]{Knowledge-grounding\\ GPT-3.5 w/o DP}} & P  & 0.800  & 0.8400 & 0.8571     \\
                                         & R  & 0.9231     & 0.8077 & 0.8182    \\
                                         & F1 & 0.8571  & 0.8235 & 0.8372    \\ \hline
\multirow{3}{*}{\shortstack[c]{Knowledge-grounding\\ GPT-3.5}} & P  & 0.8966  & \textbf{0.9565} & \textbf{1.0000}       \\
                                         & R  & \textbf{1.0000}     & 0.8462 & \textbf{0.8636}    \\ 
                                         & F1 & \textbf{0.9455}  & \textbf{0.8980} & \textbf{0.9268}    \\ \hline                                                 
\end{tabular}
\end{table}

\begin{enumerate}
    \item Unable to handle the conversion of units between data and text. For example, given the text \emph{`\texttt{The bank balance} in July 2023 is also much \texttt{lower} than \texttt{CNY 679 billion} a year earlier and \texttt{CNY 3.05 trillion} in June.'}, the numerical \texttt{CNY 3.05 trillion} was not recognized. Through checking, we found that the unit is `CNY Billion', and `CNY 3.05 trillion' corresponds to the value `3050.0'.
    \item Recognize the \texttt{numerical} that is not in the data table incorrectly. For example, given the text \emph{`\texttt{The unemployment rate} in France \texttt{inched up} to \texttt{7.2\%} in the second quarter of 2023 from \texttt{7.1\%} in the previous quarter, and the highest since Q4 2022, as the number of unemployed people increased by 19 thousand to 2.2 million.'}, the numerical \texttt{19} was incorrectly recognized, which is not in the data table.
    \item Recognize non-subject words as \texttt{subject}. For example, given the text `\emph{Lithium industry experts predict a \texttt{rising trend} in \texttt{installed wind and PV capacity (GW)}, which is expected to reach \texttt{1,200} GW and reach a \texttt{25\%} energy percentage.'}, `Lithium industry experts' was incorrectly recognized as a subject, but it is not the subject in this narrative.
\end{enumerate}

\subsection{Evaluation for \tool System}
\label{ssec:user_study}

\subsubsection{Evaluation for Graphics Overlaying Module}
\textbf{Participants.}
We recruited 18 participants (8 females, 10 males) aged 20 to 34.
Before the study, we collected participants' basic information through a questionnaire.
To demonstrate \tool is useful for users from different backgrounds, we enrolled seven experts having experience in storytelling and interactive visualization system development (E1 - E7), five novices with only experience in using visualizations in reports or courses (N1 - N5), and six users who only read and use simple financial charts (N6 - N11).

\noindent
\textbf{Setup and Procedure.}
To evaluate the \emph{graphics overlaying} module, we focused on participants' assessments of automatically generated layered charts.
We showed participants 10 sets of alternative narrative visualizations:
\begin{itemize}
    \item Side-by-side interplay without visual linking. The method is common in systems like \emph{Calliope}~\cite{calliope_2020} for data stories and \emph{AutoTitle}~\cite{autotitle_2023} for automating generating titles.
    \item Side-by-side interplay with visual linking. The method leverages visual cues like lines or colors to connect corresponding components in juxtapositioned charts and texts, adopted in systems such as \emph{CrossData}~\cite{chen2022crossdata} and \emph{DataTales}~\cite{DataTales_2023}.
    \item Layered charts with graphical overlays. The method is employed in systems like \emph{Contextifier}~\cite{hullman2013contextifier}, \emph{ChartText}~\cite{pinheiro2022charttext}, \emph{LTV}~\cite{linktv_2024}, and our \tool. In comparison to previous systems, our \tool generates more comprehensive graphical overlays dedicated for financial narrative visualizations, facilitated by the thorough examination of the correspondence between graphical overlays and financial narratives.
\end{itemize}

Specifically, the layered charts with graphical overlays were automatically generated by our system without any user refinement. Then, we manully produced the other two alternative narrative visualizations, using the same chart type and annotated texts by referring to the layered charts.
Some of the examples are presented in supplementary B.

\noindent
\textbf{Measures.}
We used a 5-point Likert scale ranging from 1 (least favorable) to 5 (most favorable), to gauge the quality of these results across three key dimensions: understandability, engagement, and comprehensiveness.
Understandability pertains to whether the layered charts and baselines are easy to understand; Engagement evaluates whether the layered charts and baselines are visually appealing and stimulating, encouraging users to explore and interact with the charts; Comprehensiveness assesses whether the charts effectively convey information behind the textual descriptions and tabular data. Participants were prompted to consider whether the charts comprehensively represented the data and vocabularies, thereby enriching the understanding of narratives.

After assigning ratings, participants were invited to provide qualitative feedback explaining their ratings. 
Additionally, they were asked to identify the chart they found clearest and the one they found least clear, shedding light on specific strengths for improvements within \tool.
These evaluations and user insights form a critical component of the comprehensive assessment of \emph{graphics overlaying} module, offering valuable feedback for its refinement and optimization.

\begin{figure}
    \centering
    \includegraphics[width = 0.99\linewidth]{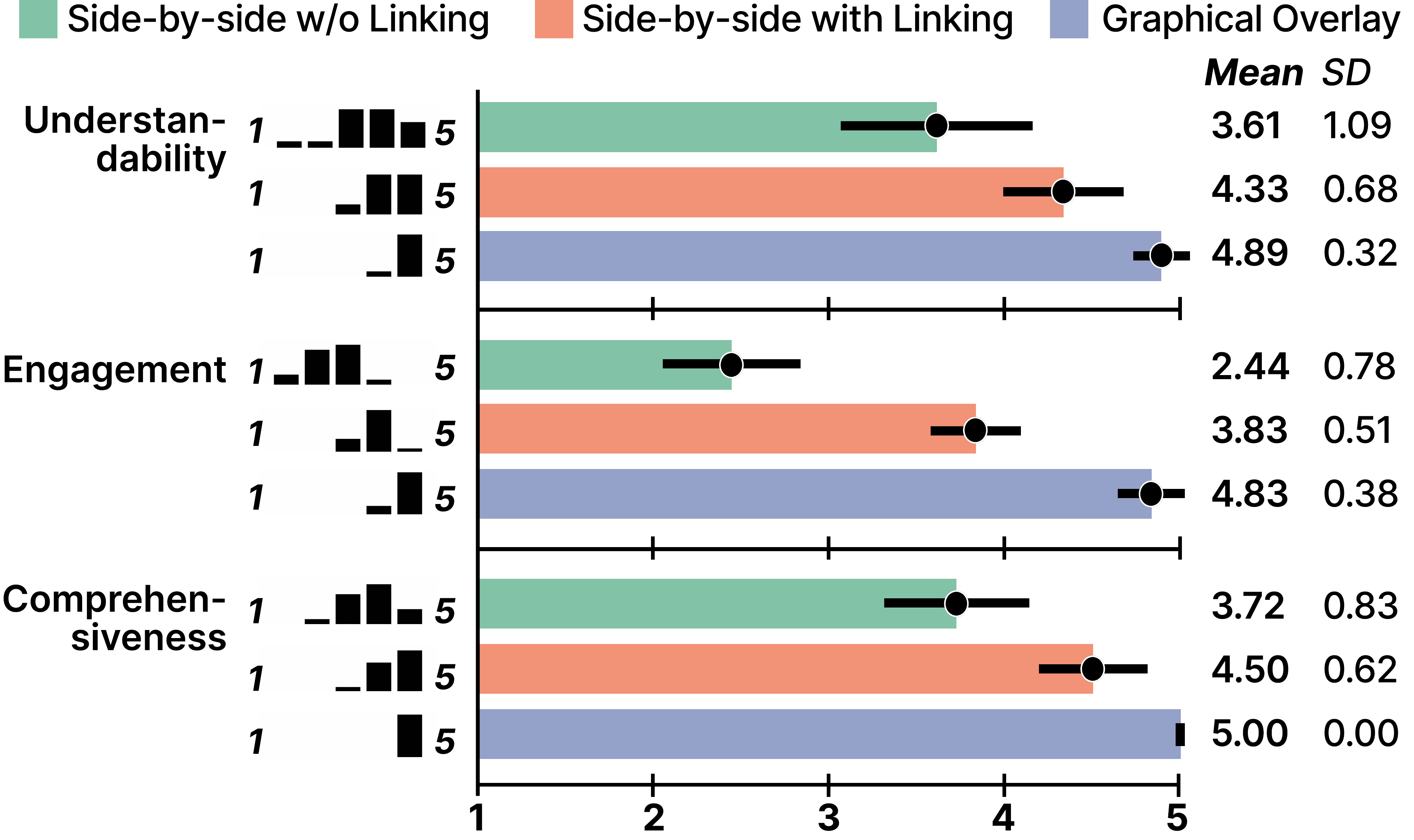}
    \vspace{-3mm}
    \caption{Ratings of the generated layered charts compared with alternative visualizations. Distributions of the ratings are shown on the left while the means and SDs are shown on the right.}
    \vspace{-4mm}
    \label{fig:errorbar}
\end{figure}

\noindent
\textbf{Results}.
Participants' ratings of the automatically generated layered charts with alternative visualizations are shown in Figure~\ref{fig:errorbar}, with error bars indicating means and standard deviations (SDs) of understandability, engagement, and comprehensiveness.
Overall, the automatically generated layered charts were preferred by participants, followed by side-by-side interplays with visual linking, and the ones without linking.
In detail, the results showed that the layered charts generated by \tool were generally well-received by the participants, with a mean score of 4.89 (16/18 participants rated it 5) for understandability, 4.83 (15/18 participants rated it 5) for engagement, and 5 (all participants rated it 5) for comprehensiveness.
For engagement, side-by-side interplay without visual linking got the lowest score, with a mean of 2.4 (SD=0.85).
Seventeen participants gave it a rating of less than 4, indicating that appropriate visual annotations would encourage users to explore narrative visualizations.

\noindent
\textbf{Findings.}
When comparing with the basic side-by-side interplay, participants praised the generated layered charts and emphasized the distinct advantages of engagement and comprehensiveness with layered charts over basic side-by-side interplay.
Especially concerning cases involving complex \texttt{trend} in horizontal analysis, participants with financial background expressed a strong preference for generated layered charts.
For instance, N6 rated the understandability of original charts with 1, because ``When reading stock market reports, the greatest challenge is pinpointing the location of the specific trend in chart, such as `triple top'. I had to examine charts featuring `resistance' and `neckline' lines to confirm the existence of this pattern. However, using \tool, I could directly get the highlighted area corresponding to `triple top', making the reading process convenient.''
In addition, participants mentioned that the default generated layered charts returned by \tool are acceptable.
``\emph{The default ones are simple but keep me focused on the key information}'' (N9).
Conversely, the layered charts are better than baselines, ``\emph{especially if the text contains long financial vocabulary}'' (N11).
The feedback indicates that the automatically generated layered chart is helpful to assist in understanding financial narratives.

\subsubsection{Evaluation for Interface and Interactions}
\textbf{Setup and Proceduce.}
The same 18 participants were also invited to participate in the evaluation for interface and interactions.
In this study, we first introduced the workflow of \tool, then instructed the participants to familiarize themselves with the system.
Finally, the participants were allowed to explore and use \tool freely. 
They could upload a data file with corresponding text, select and edit the layered charts to get satisfactory designs.
If they wanted to explore more, they could ask questions in Narrative Panel directly and got the answer with the corresponding layered chart.
Finally, the participants could export a GIF file to record text with corresponding layered charts in the narrative order.
After the exploration process, we invited the participants to rate the usability of \tool and the quality of automatically generated layered charts on a 5-point Likert scale.
After that, we conducted semi-structured one-on-one interviews with the participants to collect their comments and suggestions about our system.
The study for each participant lasted for about 40 minutes.

\noindent
\textbf{Measures.}
For the evaluation of interface and interactions, we focused on the critical aspects of usefulness, ease of use, and ease of learning.
Usefulness assesses whether \tool effectively supports participants in achieving their goals of getting appropriate layered charts.
Ease of use gauges the user-friendliness of the system's interface.
Participants rated how effortless and intuitive it was to use the system and perform their intended actions.
Ease of learning examines the \tool's accessibility to new participants. It assesses how quickly and efficiently participants believed that they could be proficient with the system and its functionalities.

\noindent
\begin{wrapfigure}[9]{r}{0.28\columnwidth}
  \vspace{-2mm}
  \includegraphics[width=0.3\columnwidth]{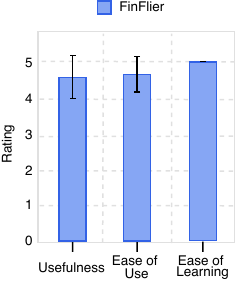}
  \label{fig:title} 
\end{wrapfigure}
\textbf{Results and Findings.}
Participants rated the system from three different criteria, with the results shown on the right.
They appreciated the usefulness (mean=4.61, SD=0.61), ease of use (mean=4.72, SD=0.46), and ease of learning (mean=5, SD=0) of \tool, illustrating that \tool offers an intuitively designed interface and interactions.
Specifically, all participants (18/18) rated 5 on ease of learning, indicating they were able to effectively use the interface after being presented with some case examples.

While experiencing the system freely, they found that ``\emph{The process of generating the default layered charts is simple, and the Design Panel is very familiar. The entire interface has no learning costs}'' (E3).
``\emph{A combination of labeling in text, highlighting in data table and layered charts could greatly complete the understanding of financial narratives.}'' (N5).
For ease of use of \tool, most participants (13/18) rated 5 on ease of use of \tool.
They expressed that it is easy for them to edit the layered charts and generate satisfactory results: ``\emph{The system is easy to get started because what needed to do are asking questions and clicking buttons}'' (E1).
Furthermore, some participants expressed valuable insights into potential aspects for system improvement.
For instance, E2 expressed a desire to support a variety of input file formats.
Some participants also suggested that the output in the form of data videos could better capture user attention.

\section{Discussion}

Reflecting on the results of the evaluation, we discuss the limitations of the current methods and explore potential directions for future research.
It also extends to broader considerations, including improving LLMs for text-data binding, as well as facilitating user engagement in narrative visualizations.

\subsection{Limitations and Opportunities}\label{ssec:limit}

\noindent
\textbf{Layered Chart Corpus.}
Our layered chart corpus integrates sources from both practical applications and academic research.
It offers a comprehensive collection, with a balanced representation of layered charts from diverse sources, supporting the correspondence between graphical overlays and financial narratives.
To further enhance its comprehensiveness and diversity, future expansions could incorporate additional sources such as Yahoo Finance, Wind, and Bloomberg.
For instance, Bloomberg's insight articles\footnote{https://www.bloomberg.com/professional/insights/} often feature black-background layered charts generated through Bloomberg Terminal.
Nevertheless, integrating new sources will demand significant manual annotation efforts.
This labor-intensive process introduces inherent limitations, as some layered charts may have been inadvertently excluded, potentially leading to biases in chart distribution~\cite{visatlas_2024}.
To address these challenges, the development of auto-labeling tools and collaborative contributions from the visualization community will be essential.

\noindent
\textbf{Expanding Chart Types and Graphical Overlays.}
The current work focuses on four basic chart types that cover a substantial portion of analysis scenarios within financial narrative visualizations, as most financial narratives are related to time-series data, which is effectively displayed by the four chart types.
However, certain cases may necessitate a broader range of chart types to achieve optimal narrative effects.
We anticipate expanding chart types, including pie charts, scatterplots, and others.
Simultaneously, we intend to expand our survey into graphical overlays for different charts, such as colored regions or partition lines in scatterplots, to facilitate the comprehension of clustering patterns.

This also presents new opportunities for a comprehensive understanding of graphical overlay techniques. 
A promising avenue for future research is the systematic exploration of the advantages and limitations of each type or combination of graphical overlays, particularly when applied across diverse scenarios.
For instance, combining multiple overlays might enhance data comprehension but potentially introduce visual clutter, thereby reducing overall interpretability. 
Similarly, adding \emph{Trend-line} might aid in locating trends in data but may be less effective in high-density multi-line charts.
A deeper examination of these trade-offs could inform guidelines for choosing appropriate overlay techniques tailored to specific tasks, user expertise, or visualization complexity.

\noindent
\textbf{Expanding Financial Narratives and Data Tables.}
The current system only supports the input of a single data file along with its corresponding textual content.
However, data-rich financial documents often consist of multiple complex data tables and long textual content.
During the user study, E2 expressed that ``it would be convenient if I could upload a file with multiple tables and textual contents, and the system generates layered charts for me directly.''
The functions can be found in Elastic Documents~\cite{badam_2019_elastic}, which supports a variety of input file formats and offers multiple candidate data tables to help users understand the full text.
Nevertheless, enabling more data tables and longer narratives is more challenging, due to the limitation posed by the token limit of LLMs.
In the following studies, we aim to expand the system's capabilities to accept complete financial documents as inputs by leveraging fine-tuned LLMs or training an LLM with a localized financial knowledge base.

\begin{figure}[t]
    \centering
    \includegraphics[width = 0.95\linewidth]{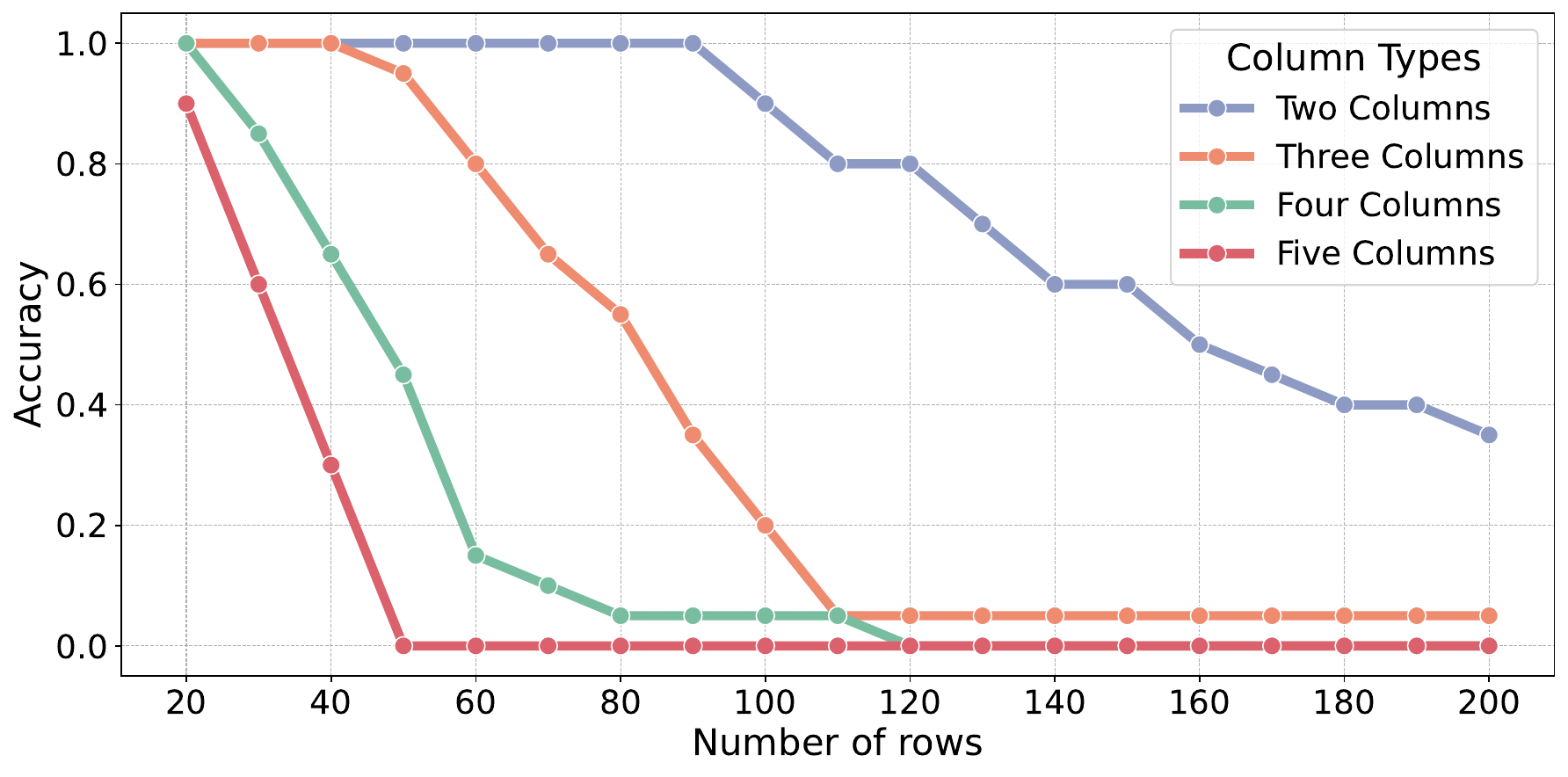}
    \vspace{-4mm}
    \caption{The accuracy of our knowledge-grounding LLM with different input table size.}
    \vspace{-3mm}
    \label{fig:scale}
\end{figure}

Besides, to test how their sizes affect the accuracy of our model, we conducted controlled experiments with varying input sizes.
We started with a data table containing 2 columns and 20 rows, accompanied by textual content with a \texttt{numerical} value.
Next, to evaluate the impact of table size, we randomly added new rows and columns to the original tables while keeping the text length constant.
Figure~\ref{fig:scale} depicts the results between the input table size and the accuracy of our method, indicating a decrease in accuracy as the table size increases.
Notably, increasing the number of columns has a greater impact than increasing the number of rows, leading to difficulties in locating data within the tales.
The results are consistent with existing studies revealing that the LLMs face difficulties in dealing with large tables ~\cite{chain_of_table}.
To assess the impact of text length, we add irrelevant text to the beginning and end of the original textual content while keeping the data tables unchanged.
We did not observe an obvious decrease on accuracy when text size increases, likely due to the powerful text-processing capabilities of LLMs.
In future work, to apply \tool to more data-rich scenarios, we plan to incorporate program-aided methods (\eg,~\cite{dater_2023}) or new tabular representations (\eg,~\cite{visltr_2024}) to achieve robust text-data binding.

\subsection{Improving LLMs for Text-data Binding}

LLMs are powerful and versatile, offering significant flexibility in handling ambiguous text compared to traditional NLP methods~\cite{logic_2020}.
Our work capitalizes on the advantages of LLMs to achieve impressive results with few prompt examples and optimization methods.
Despite this, issues such as mismatches in text-data binding and hallucination persist.
These challenges occasionally lead to errors in generated responses, impacting the precision of text-data binding and subsequent graphics overlaying.
To mitigate the problem, we have implemented a feedback loop in the current pipeline.
Users can provide feedback on system-generated responses, and if they find the results unsatisfactory, they can click on the Thumb-down button to request a regenerated response.
However, this iterative refinement mechanism cannot completely eliminate inaccuracies or omissions.
Addressing open-ended questions and unsummarized patterns may exacerbate hallucination risks~\cite{hall_survey}.
To extend \tool's capabilities to provide more accurate and context-aware responses in financial narrative visualizations, promising strategies include refining LLMs through domain-specific fine-tuning, incorporating external knowledge bases~\cite{rag_hall}, and leveraging multi-agent debate mechanisms~\cite{multiagent_hall} to validate reasoning and enhance reliability.
Future iterations of \tool could explore integrating such methods to further enhance the quality and relevance of responses in more general financial narratives.

This work has distilled three types of vocabularies: \texttt{subject}, \texttt{trend}, and \texttt{numerical} in financial domain.
During interviews, N1 and N7 pointed out that ``the text-data binding in Narrative Panel played a significant role in the understanding of narrative structures, especially long complex vocabularies.''
It suggests that our knowledge-grounding LLM-based \emph{text-data binding} module has the potential to extend its applicability beyond the field of finance.
For instance, in the medical domain, with a few prompt examples, it could assist in identifying and labeling lengthy, domain-specific terms.
When encountering content that users cannot understand, users can directly pose natural language questions and receive responses in Narrative Panel.
This adaptability positions our \emph{text-data binding} module as a versatile tool with broader applicability across diverse knowledge domains.

\subsection{Facilitating User Engagement in Narrative Visualization}
This work focuses on graphical overlays which are effective in conveying insights and context in financial narrative visualizations.
Although the automatically generated layered charts were preferred by participants over baselines in user studies, five participants (E1, E4, N2, N4, and N10) noted that ``compared with static layered charts, there is potential for data videos to further improve the comprehension and engagement of complex narratives.''
They suggested that incorporating animation can be beneficial, allowing for the removal of unnecessary visual elements or the merging of related elements, which may facilitate user engagement compared to static visualizations.
Our system lays a solid foundation for exploring data videos in financial domain.
The following research may include animations, videos, or interactive elements that can be seamlessly integrated with \tool.
We anticipate that animation techniques will be particularly valuable when addressing complex analyses and engaging diverse audiences, thus requiring further exploration in future research.

\section{conclusion}
In this paper, we present a novel system, \tool, which leverages a knowledge-grounding LLM to automate graphical overlays for financial visualizations.
Based on the summary of commonly used graphical overlays and financial narrative structures, we design two core modules for the system, namely \emph{text-data binding} and \emph{graphics overlaying}.
The \emph{text-data binding} module enhances the LLM-based connection between financial vocabulary and data table through advanced prompt engineering techniques, including output constraint, CoT and dynamic prompt.
The \emph{graphics overlaying} module generates effective layered charts, considering narrative sequencing and correspondence between graphical overlays and financial narratives.
We further develop an interactive visual interface that supports natural language interactions and allows for the flexible configuration of graphical overlays.
Users can seamlessly leverage the system to quickly explore financial narrative visualizations, and create and iterate on the corresponding layered charts.
Furthermore, we conduct case studies covering three types of financial analysis: \emph{horizontal analysis}, \emph{vertical analysis} and \emph{combined analysis} to demonstrate the applicability of \tool to financial narrative visualizations.
Feedback and ratings from users confirm the feasibility and effectiveness of \tool and high quality of automatically generated layered charts.
Our open-sourced corpus, code, and examples would promote future research for financial narrative visualizations and graphical overlays.

 \section*{Acknowledgements}
 This work is supported/funded by the Guangzhou-HKUST(GZ) Joint Funding Program (No. 2024A03J0630)

\bibliographystyle{IEEEtran}
\bibliography{template}

\begin{thebibliography}{10}
\providecommand{\url}[1]{#1}
\csname url@samestyle\endcsname
\providecommand{\newblock}{\relax}
\providecommand{\bibinfo}[2]{#2}
\providecommand{\BIBentrySTDinterwordspacing}{\spaceskip=0pt\relax}
\providecommand{\BIBentryALTinterwordstretchfactor}{4}
\providecommand{\BIBentryALTinterwordspacing}{\spaceskip=\fontdimen2\font plus
\BIBentryALTinterwordstretchfactor\fontdimen3\font minus \fontdimen4\font\relax}
\providecommand{\BIBforeignlanguage}[2]{{%
\expandafter\ifx\csname l@#1\endcsname\relax
\typeout{** WARNING: IEEEtran.bst: No hyphenation pattern has been}%
\typeout{** loaded for the language `#1'. Using the pattern for}%
\typeout{** the default language instead.}%
\else
\language=\csname l@#1\endcsname
\fi
#2}}
\providecommand{\BIBdecl}{\relax}
\BIBdecl

\bibitem{tellingfinance_2000}
D.~A. Jameson, ``{Telling the investment story}: A narrative analysis of shareholder reports,'' \emph{The Journal of Business Communication}, vol.~37, no.~1, pp. 7--38, 2000.

\bibitem{narrineco_2024}
M.~Roos and M.~Reccius, ``{N}arratives in economics,'' \emph{Journal of Economic Surveys}, vol.~38, no.~2, pp. 303--341, 2024.

\bibitem{narrvis_2010}
E.~Segel and J.~Heer, ``{N}arrative visualization: Telling stories with data,'' \emph{IEEE Trans. Vis. Comput. Graph.}, vol.~16, no.~6, pp. 1139--1148, 2010.

\bibitem{lai2020automatic}
C.~Lai, Z.~Lin, R.~Jiang, Y.~Han, C.~Liu, and X.~Yuan, ``{A}utomatic {A}nnotation {S}ynchronizing with {T}extual {D}escription for {V}isualization,'' in \emph{Proc. ACM CHI}, 2020, pp. 1--13.

\bibitem{pinheiro2022charttext}
J.~Pinheiro and J.~Poco, ``{ChartText}: Linking text with charts in documents,'' \emph{arXiv preprint arXiv:2201.05043}, 2022.

\bibitem{chen2022crossdata}
Z.~Chen and H.~Xia, ``{CrossData}: Leveraging text-data connections for authoring data documents,'' in \emph{Proc. ACM CHI}, 2022, pp. 1--15.

\bibitem{masson2023charagraph}
D.~Masson, S.~Malacria, G.~Casiez, and D.~Vogel, ``{Charagraph}: Interactive generation of charts for realtime annotation of data-rich paragraphs,'' in \emph{Proc. ACM CHI}, 2023, pp. 1--18.

\bibitem{DataTales_2023}
N.~Sultanum and A.~Srinivasan, ``{DATATALES}: Investigating the use of large language models for authoring data-driven articles,'' in \emph{Proc. IEEE VIS}, 2023, pp. 231--235.

\bibitem{graphicalo_2012}
N.~Kong and M.~Agrawala, ``Graphical overlays: Using layered elements to aid chart reading,'' \emph{IEEE Trans. Vis. Comput. Graph.}, vol.~18, no.~12, pp. 2631--2638, 2012.

\bibitem{hullman2013contextifier}
J.~Hullman, N.~Diakopoulos, and E.~Adar, ``{Contextifier}: Automatic generation of annotated stock visualizations,'' in \emph{Proc. ACM CHI}, 2013, pp. 2707--2716.

\bibitem{srinivasan2018augmenting}
A.~Srinivasan, S.~M. Drucker, A.~Endert, and J.~Stasko, ``Augmenting visualizations with interactive data facts to facilitate interpretation and communication,'' \emph{IEEE Trans. Vis. Comput. Graph.}, vol.~25, no.~1, pp. 672--681, 2018.

\bibitem{godesign_2023}
M.~D. Rahman, G.~J. Quadri, B.~Doppalapudi, D.~A. Szafir, and P.~Rosen, ``{A} {Q}ualitative {A}nalysis of {C}ommon {P}ractices in {A}nnotations: A taxonomy and design space,'' \emph{arXiv preprint arXiv:2306.06043}, 2023.

\bibitem{finance_2004}
P.~Castells, B.~Foncillas, R.~Lara, M.~Rico, and J.~L. Alonso, ``Semantic web technologies for economic and financial information management,'' in \emph{Proc. ESWS}, 2004, pp. 473--487.

\bibitem{visltr_2024}
J.~Hao, Z.~Liang, C.~Li, Y.~Luo, j.~Li, and W.~Zeng, ``{VisTR}: Visualizations as representations for time-series table reasoning,'' \emph{arXiv preprint arXiv:2406.03753}, 2024.

\bibitem{zhou2023one}
T.~Zhou, P.~Niu, X.~Wang, L.~Sun, and R.~Jin, ``{One Fits All}: Power general time series analysis by pretrained lm,'' \emph{arXiv preprint arXiv:2302.11939}, 2023.

\bibitem{linktv_2024}
X.~Cai, D.~Weng, T.~Fu, S.~Fu, Y.~Wang, and Y.~Wu, ``Linking text and visualizations via contextual knowledge graph,'' \emph{IEEE Trans. Vis. Comput. Graph.}, pp. 1--14, 2024.

\bibitem{segel2010narrative}
E.~Segel and J.~Heer, ``{Narrative Visualization}: Telling stories with data,'' \emph{IEEE Trans. Vis. Comput. Graph.}, vol.~16, no.~6, pp. 1139--1148, 2010.

\bibitem{calliope_2020}
D.~Shi, X.~Xu, F.~Sun, Y.~Shi, and N.~Cao, ``{Calliope}: Automatic visual data story generation from a spreadsheet,'' \emph{IEEE Trans. Vis. Comput. Graph.}, vol.~27, no.~2, pp. 453--463, 2020.

\bibitem{dataplayer_2023}
L.~Shen, Y.~Zhang, H.~Zhang, and Y.~Wang, ``{Data Player}: Automatic generation of data videos with narration-animation interplay,'' \emph{IEEE Trans. Vis. Comput. Graph.}, vol.~30, no.~1, pp. 109--119, 2024.

\bibitem{shi2018deepclue}
L.~Shi, Z.~Teng, L.~Wang, Y.~Zhang, and A.~Binder, ``{DeepClue}: Visual interpretation of text-based deep stock prediction,'' \emph{IEEE Transactions on Knowledge and Data Engineering}, vol.~31, no.~6, pp. 1094--1108, 2018.

\bibitem{ko_2016_survey}
S.~Ko, I.~Cho, S.~Afzal, C.~Yau, J.~Chae, A.~Malik, K.~Beck, Y.~Jang, W.~Ribarsky, and D.~S. Ebert, ``A survey on visual analysis approaches for financial data,'' \emph{Comput. Graph. Forum}, vol.~35, no.~3, pp. 599--617, 2016.

\bibitem{koop2022analysis}
G.~Koop, \emph{Analysis of financial data}.\hskip 1em plus 0.5em minus 0.4em\relax John Wiley \& Sons, Ltd, 2006.

\bibitem{brown2020language}
T.~Brown, B.~Mann, N.~Ryder, M.~Subbiah, J.~D. Kaplan, P.~Dhariwal, A.~Neelakantan, P.~Shyam, G.~Sastry, A.~Askell \emph{et~al.}, ``Language models are few-shot learners,'' \emph{Adv. Neural Inf. Process.}, vol.~33, pp. 1877--1901, 2020.

\bibitem{latif2021kori}
S.~Latif, Z.~Zhou, Y.~Kim, F.~Beck, and N.~W. Kim, ``{Kori}: Interactive synthesis of text and charts in data documents,'' \emph{IEEE Trans. Vis. Comput. Graph.}, vol.~28, no.~1, pp. 184--194, 2021.

\bibitem{wordsize_2017}
F.~Beck and D.~Weiskopf, ``{W}ord-sized graphics for scientific texts,'' \emph{IEEE Trans. Vis. Comput. Graph.}, vol.~23, no.~6, pp. 1576--1587, 2017.

\bibitem{storytelling_2013}
R.~Kosara and J.~Mackinlay, ``{Storytelling}: The next step for visualization,'' \emph{Comput.}, vol.~46, no.~5, pp. 44--50, 2013.

\bibitem{ren2017chartaccent}
D.~Ren, M.~Brehmer, B.~Lee, T.~H{\"o}llerer, and E.~K. Choe, ``{ChartAccent}: Annotation for data-driven storytelling,'' in \emph{Proc. IEEE PacificVis}, 2017, pp. 230--239.

\bibitem{zhi2019linking}
Q.~Zhi, A.~Ottley, and R.~Metoyer, ``Linking and layout: Exploring the integration of text and visualization in storytelling,'' \emph{Comput. Graph. Forum.}, vol.~38, no.~3, pp. 675--685, 2019.

\bibitem{badam_2019_elastic}
S.~K. Badam, Z.~Liu, and N.~Elmqvist, ``{Elastic Documents}: Coupling text and tables through contextual visualizations for enhanced document reading,'' \emph{IEEE Trans. Vis. Comput. Graph.}, vol.~25, no.~1, pp. 661--671, 2019.

\bibitem{head2021augmenting}
A.~Head, K.~Lo, D.~Kang, R.~Fok, S.~Skjonsberg, D.~S. Weld, and M.~A. Hearst, ``Augmenting scientific papers with just-in-time, position-sensitive definitions of terms and symbols,'' in \emph{Proc. ACM CHI}, 2021, pp. 1--18.

\bibitem{kong2014extracting}
N.~Kong, M.~A. Hearst, and M.~Agrawala, ``Extracting references between text and charts via crowdsourcing,'' in \emph{Proc. ACM CHI}, 2014, pp. 31--40.

\bibitem{stokes2022striking}
C.~Stokes, V.~Setlur, B.~Cogley, A.~Satyanarayan, and M.~A. Hearst, ``Striking a balance: Reader takeaways and preferences when integrating text and charts,'' \emph{IEEE Trans. Vis. Comput. Graph.}, vol.~29, no.~1, pp. 1233--1243, 2022.

\bibitem{latif2018exploring}
S.~Latif, D.~Liu, and F.~Beck, ``Exploring interactive linking between text and visualization.'' in \emph{Proc. EuroVis}, 2018, pp. 91--94.

\bibitem{kong2017internal}
H.-K. Kong, Z.~Liu, and K.~Karahalios, ``Internal and external visual cue preferences for visualizations in presentations,'' \emph{Comput. Graph. Forum}, vol.~36, no.~3, pp. 515--525, 2017.

\bibitem{emphchecker_2023}
D.~H. Kim, S.~Choi, J.~Kim, V.~Setlur, and M.~Agrawala, ``{EmphasisChecker}: A tool for guiding chart and caption emphasis,'' \emph{IEEE Trans. Vis. Comput. Graph.}, vol.~30, no.~1, pp. 120--130, 2024.

\bibitem{kim2018facilitating}
D.~H. Kim, E.~Hoque, J.~Kim, and M.~Agrawala, ``{F}acilitating {D}ocument {R}eading by {L}inking {T}ext and {T}ables,'' in \emph{Proc. ACM UIST}, 2018, pp. 423--434.

\bibitem{difference_2023}
D.~Bromley and V.~Setlur, ``{What Is the Difference Between a Mountain and a Molehill?} quantifying semantic labeling of visual features in line charts,'' in \emph{Proc. IEEE VIS}, 2023, pp. 161--165.

\bibitem{liu2022few}
H.~Liu, D.~Tam, M.~Muqeeth, J.~Mohta, T.~Huang, M.~Bansal, and C.~A. Raffel, ``Few-shot parameter-efficient fine-tuning is better and cheaper than in-context learning,'' \emph{Adv. Neural Inf. Process.}, vol.~35, pp. 1950--1965, 2022.

\bibitem{cot_2022}
J.~Wei, X.~Wang, D.~Schuurmans, M.~Bosma, F.~Xia, E.~Chi, Q.~V. Le, D.~Zhou \emph{et~al.}, ``Chain-of-thought prompting elicits reasoning in large language models,'' \emph{Adv. Neural Inf. Process.}, vol.~35, pp. 24\,824--24\,837, 2022.

\bibitem{ouyang2022training}
L.~Ouyang, J.~Wu, X.~Jiang, D.~Almeida, C.~Wainwright, P.~Mishkin, C.~Zhang, S.~Agarwal, K.~Slama, A.~Ray \emph{et~al.}, ``Training language models to follow instructions with human feedback,'' \emph{Adv. Neural Inf. Process.}, vol.~35, pp. 27\,730--27\,744, 2022.

\bibitem{huang2023finbert}
A.~H. Huang, H.~Wang, and Y.~Yang, ``{FinBERT}: A large language model for extracting information from financial text,'' \emph{Contemporary Accounting Research}, vol.~40, no.~2, pp. 806--841, 2023.

\bibitem{yang2023fingpt}
H.~Yang, X.-Y. Liu, and C.~D. Wang, ``{FinGPT}: Open-source financial large language models,'' \emph{arXiv preprint arXiv:2306.06031}, 2023.

\bibitem{GAIsur_2024}
Y.~Ye, J.~Hao, Y.~Hou, Z.~Wang, S.~Xiao, Y.~Luo, and W.~Zeng, ``Generative {AI} for visualization: State of the art and future directions,'' \emph{Visual Informatics}, vol.~8, no.~2, pp. 43--66, 2024.

\bibitem{chain_of_table}
Z.~Wang, H.~Zhang, C.-L. Li, J.~M. Eisenschlos, V.~Perot, Z.~Wang, L.~Miculicich, Y.~Fujii, J.~Shang, C.-Y. Lee \emph{et~al.}, ``{C}hain-of-{T}able: {E}volving {T}ables in the {R}easoning {C}hain for {T}able {U}nderstanding,'' \emph{arXiv preprint arXiv:2401.04398}, 2024.

\bibitem{hyland_1998}
K.~Hyland, \emph{Hedging in scientific research articles}.\hskip 1em plus 0.5em minus 0.4em\relax John Benjamins Publishing Company, 1998.

\bibitem{ravinderFinancialAnalysisStudy2013}
D.~Ravinder, ``Financial analysis – a study,'' \emph{IOSR Journal of Economics and Finance}, vol.~2, pp. 10--22, 2013.

\bibitem{aichain_2022}
T.~Wu, M.~Terry, and C.~J. Cai, ``{AI Chains}: Transparent and controllable human-ai interaction by chaining large language model prompts,'' in \emph{Proc. ACM CHI}, 2022, pp. 1--22.

\bibitem{aiprom_2021}
J.~O'Connor and J.~Andreas, ``What context features can transformer language models use?'' \emph{arXiv preprint arXiv:2106.08367}, 2021.

\bibitem{structureoutput_2024}
M.~X. Liu, F.~Liu, A.~J. Fiannaca, T.~Koo, L.~Dixon, M.~Terry, and C.~J. Cai, ``{``We Need Structured Output"}: Towards user-centered constraints on large language model output,'' in \emph{Proceedings of Extended Abstracts of the CHI Conference on Human Factors in Computing Systems}, 2024, pp. 1--9.

\bibitem{logic_2021}
G.~Betz, K.~Richardson, and C.~Voigt, ``{T}hinking aloud: {D}ynamic context generation improves zero-shot reasoning performance of gpt-2,'' \emph{arXiv preprint arXiv:2103.13033}, 2021.

\bibitem{logic_2020}
L.~Floridi and M.~Chiriatti, ``{GPT-3:} {I}ts {N}ature, {S}cope, {L}imits, and {C}onsequences,'' \emph{Minds Mach.}, vol.~30, no.~4, pp. 681--694, 2020.

\bibitem{weakllm_2020}
E.~M. Bender and A.~Koller, ``{C}limbing towards {NLU}: On meaning, form, and understanding in the age of data,'' in \emph{Proc. ACL}, 2020, pp. 5185--5198.

\bibitem{prompteng_2023}
G.~Marvin, N.~Hellen, D.~Jjingo, and J.~Nakatumba-Nabende, ``Prompt engineering in large language models,'' in \emph{Proc. ICDICI}, 2023, pp. 387--402.

\bibitem{LM-BFF_2020}
T.~Gao, A.~Fisch, and D.~Chen, ``Making pre-trained language models better few-shot learners,'' \emph{arXiv preprint arXiv:2012.15723}, 2020.

\bibitem{gestalt_1938}
M.~Wertheimer, ``Laws of organization in perceptual forms.'' \emph{Psycologische Forschung}, vol.~4, 1923.

\bibitem{hall_survey}
Z.~Ji, N.~Lee, R.~Frieske, T.~Yu, D.~Su, Y.~Xu, E.~Ishii, Y.~J. Bang, A.~Madotto, and P.~Fung, ``Survey of hallucination in natural language generation,'' \emph{ACM Comput. Surv.}, vol.~55, no.~12, pp. 1--38, 2023.

\bibitem{autotitle_2023}
C.~Liu, Y.~Guo, and X.~Yuan, ``{AutoTitle}: An interactive title generator for visualizations,'' \emph{IEEE Trans. Vis. Comput. Graph.}, vol.~30, no.~8, pp. 5276 -- 5288, 2023.

\bibitem{visatlas_2024}
Y.~Ye, R.~Huang, and W.~Zeng, ``{VISAtlas}: An image-based exploration and query system for large visualization collections via neural image embedding,'' \emph{IEEE Trans. Vis. Comput. Graph.}, vol.~30, no.~7, pp. 3224--3240, 2024.

\bibitem{dater_2023}
Y.~Ye, B.~Hui, M.~Yang, B.~Li, F.~Huang, and Y.~Li, ``{L}arge language models are versatile decomposers: {D}ecomposing evidence and questions for table-based reasoning,'' in \emph{Proc. ACM SIGIR}, 2023, pp. 174--184.

\bibitem{rag_hall}
J.~Li, Y.~Yuan, and Z.~Zhang, ``Enhancing llm factual accuracy with rag to counter hallucinations: {A} case study on domain-specific queries in private knowledge-bases,'' \emph{arXiv preprint arXiv:2403.10446}, 2024.

\bibitem{multiagent_hall}
J.~Zhu, P.~Cai, K.~Xu, L.~Li, Y.~Sun, S.~Zhou, H.~Su, L.~Tang, and Q.~Liu, ``{AutoTQA: Towards Autonomous Tabular Question Answering through Multi-Agent Large Language Models},'' \emph{Proc. {VLDB} Endow.}, vol.~17, no.~12, pp. 3920--3933, 2024.

\end{thebibliography}

\vspace{-33pt}
\begin{IEEEbiography}[{\includegraphics[width=1in,height=1.25in,clip,keepaspectratio]{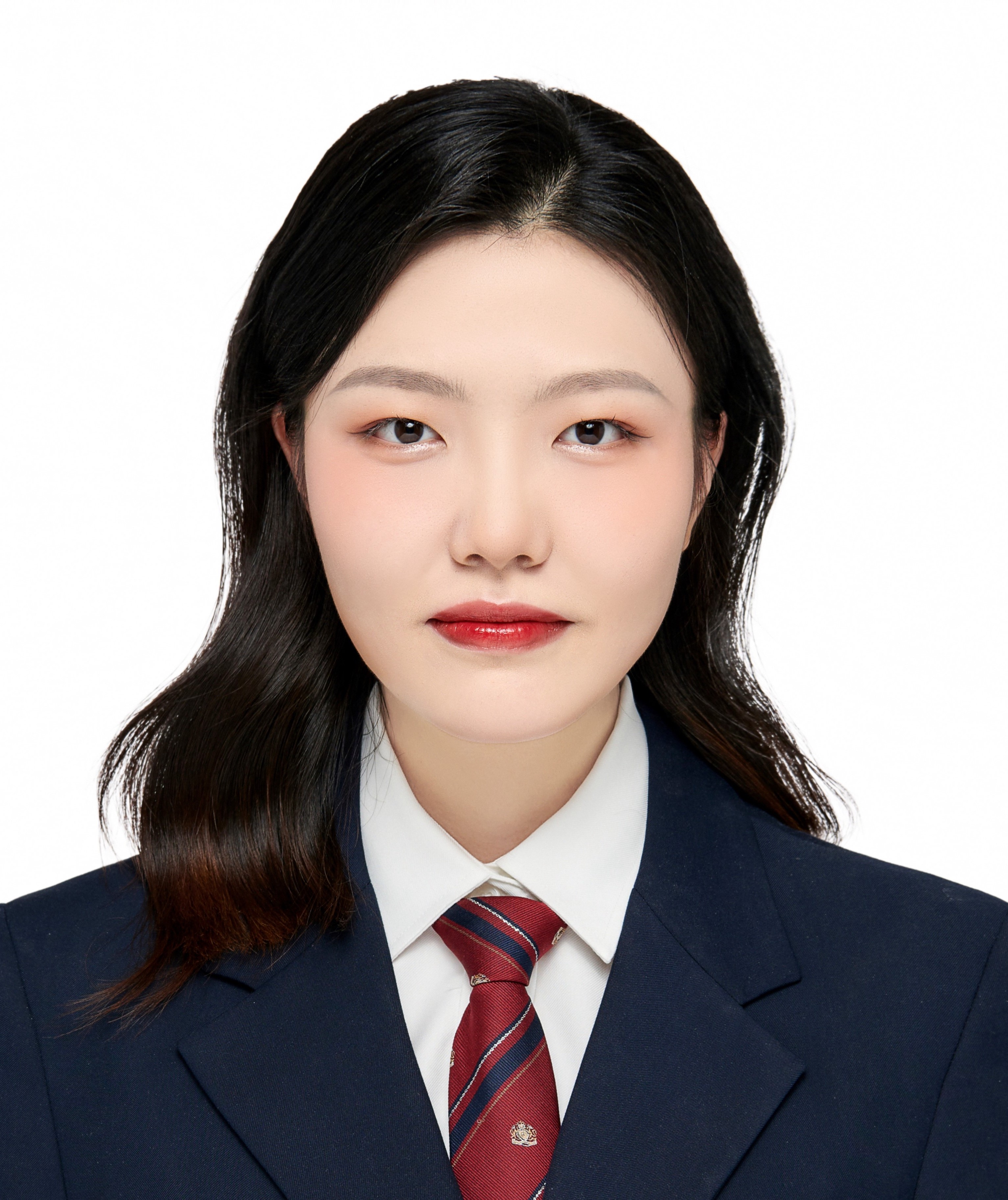}}]{Jianing Hao}
received a B.E. degree in computer science and technology from Shandong University, China, in 2022. She is currently a Ph.D. student at The Hong Kong University of Science and Technology (Guangzhou). Her research interests include time-series data, Human-AI collaboration, and visual analytics.
\end{IEEEbiography}

\vspace{-30pt}

\begin{IEEEbiography}[{\includegraphics[width=1in,height=1.25in,clip,keepaspectratio]{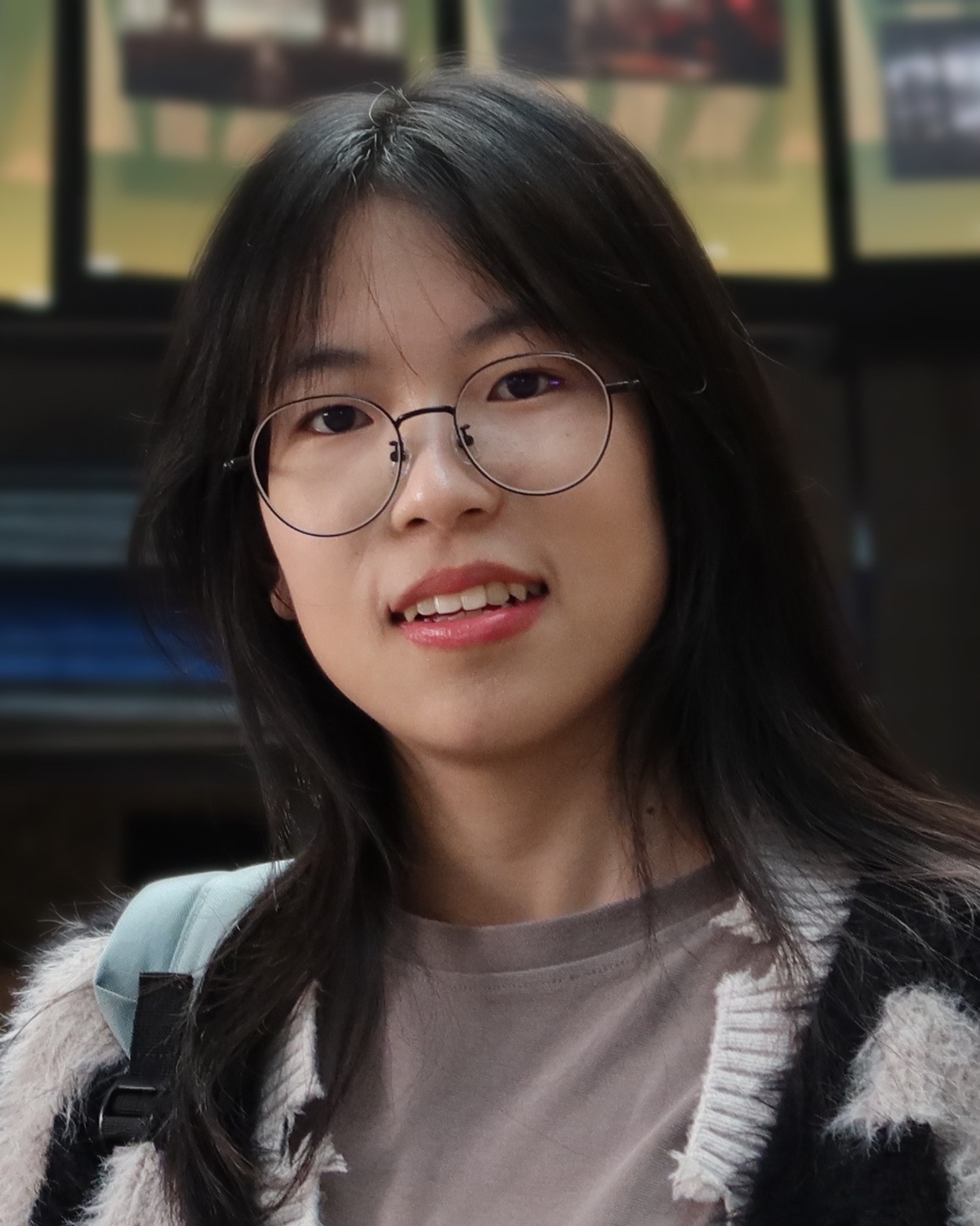}}]{Manling Yang} is an MPhil student at the Hong Kong University of Science and Technology (Guangzhou). Her research interests include data visualization and human-computer interaction. She received her B.E. in Electronic Information Engineering from Chongqing University, China in 2023.
\end{IEEEbiography}

\vspace{-30pt}

\begin{IEEEbiography}[{\includegraphics[width=1in,height=1.25in,clip,keepaspectratio]{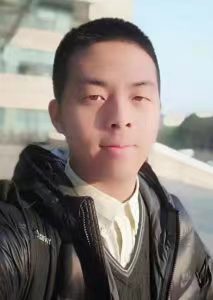}}]
{Qing Shi} was a research assistant at the Computational Media and Arts (CMA) Thrust of the Hong Kong University of Science and
Technology (Guangzhou). His recent research interests include visualization and visual analytics, XAI, and HCI.
\end{IEEEbiography}

\vspace{-30pt}

\begin{IEEEbiography}[{\includegraphics[width=1in,height=1.25in,clip,keepaspectratio]{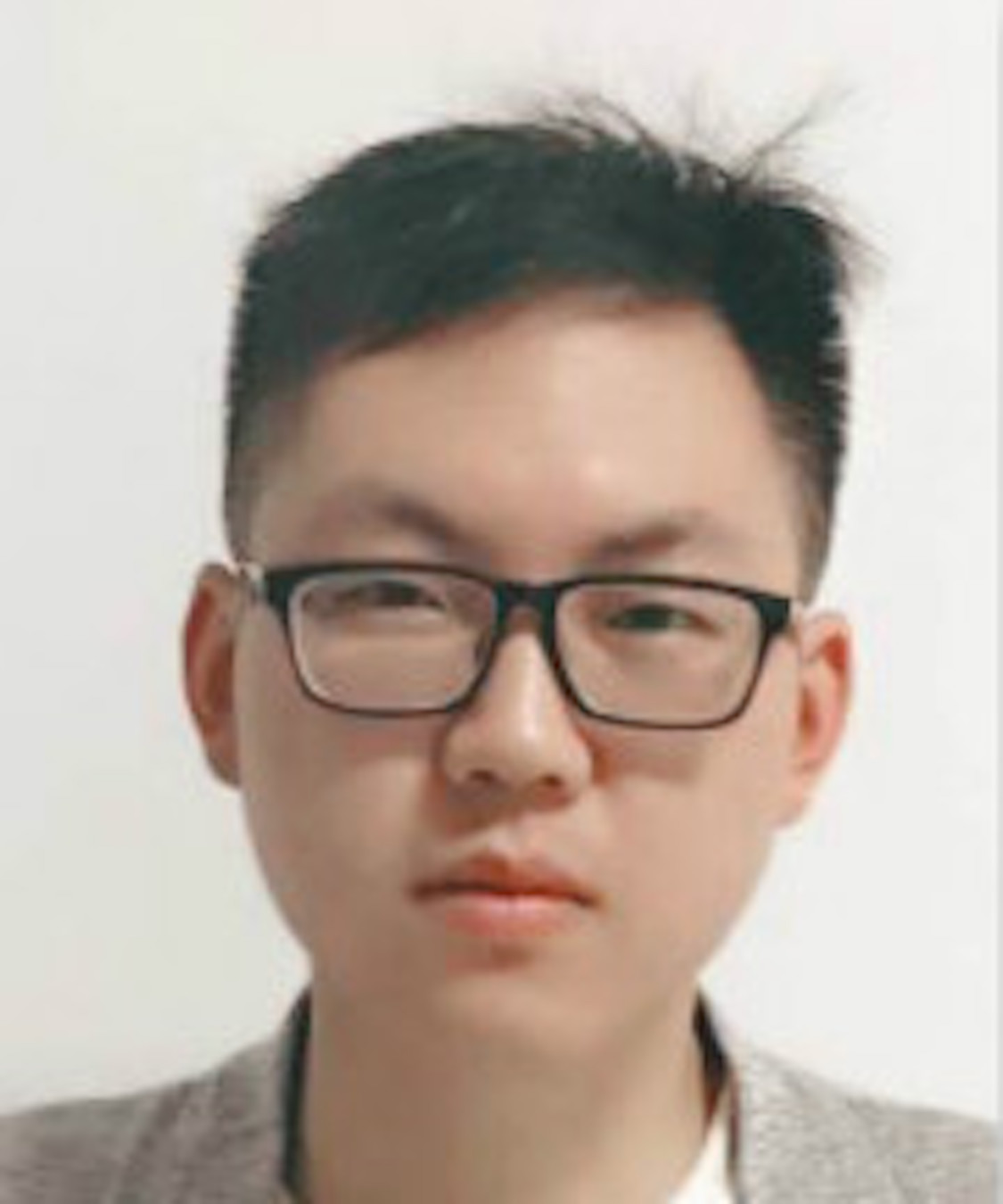}}]{Yuzhe Jiang}
is currently a Ph.D. candidate in the Department of Computer Science and Engineering at the Hong Kong University of Science
and Technology (HKUST). He received his B.E. degree in Computer Science and Technology from University of Chinese Academy of Sciences, China in 2018. His research interests include visualization and Human-AI collaboration.
\end{IEEEbiography}

\vspace{-30pt}

\begin{IEEEbiography}[{\includegraphics[width=1in,height=1.25in,clip,keepaspectratio]{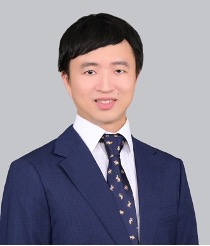}}]{Guang Zhang}
Guang Zhang is an assistant professor at the Hong Kong University of Science and Technology (Guangzhou), and affiliated assistant professor at the Hong Kong University of Science and Technology. He received his Ph.D. degree in economics from Boston University in 2021. His research interests include FinTech, financial econometrics, empirical finance and machine learning.
\end{IEEEbiography}

\vspace{-30pt}

\begin{IEEEbiography}[{\includegraphics[width=1.0in,height=1.25in,clip,keepaspectratio]{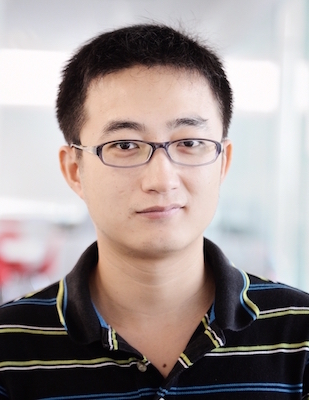}}]
{Wei Zeng} is an assistant professor at the Hong Kong University of Science and Technology (Guangzhou). He received his Ph.D. in computer science from Nanyang Technological University in 2015. He received Best Paper Awards from ICIV, VINCI, and ChinaVis. He served as Program Chair for VINCI’23, and program committee for venues including IEEE VIS, EuroVis STARs, and ChinaVis. His recent research interests include visualization and visual analytics, and AIGC.
\end{IEEEbiography}

\vfill
\newpage

\end{document}